\documentclass[%
aps,12pt,draft,
 prd,
 superscriptaddress,
 amsmath,amssymb,
 aps,nofootinbib,   
]{revtex4}

\usepackage{array}
\usepackage{mathtools}

\usepackage{etoolbox}

\begin{document} 
\title{Cosmology as a Crossed Product}

\author{C\'esar G\'omez} 
\affiliation{Instituto de F\'{i}sica Te\'orica UAM-CSIC, Universidad Aut\'onoma de Madrid, Cantoblanco, 28049 Madrid, Spain.}




\date{\today}


\begin{abstract}
We suggest a new conceptual frame where inflationary Cosmology is quantum mechanically described using the Hilbert space representation of the crossed product of the type $III$ factor associated with the algebra of local operators on the de Sitter static patch and the automorphism of translations in the emergent conformal time. In this formal framework scalar curvature fluctuations are determined by the gravitational potential induced by the quantum variance of the generator of the conformal time automorphisms. Choosing as representative, in the Hilbert space of the crossed product, the effective vacuum state for each mode and conformal time, we evaluate the quantum variance and the induced scalar curvature fluctuations. This leads to a model independent characterization of the inflationary parameters $(1-n_s) \sim 0.0318$ and $\epsilon\sim 0.0027$ in good agreement with Planck experimental results. In this formal frame inflation extends the algebra of observables in a primordial dS expanding Universe. In this extension time evolution becomes an inner automorphism and the conjugated energy fluctuations manifest as the observable scalar curvature fluctuations.

\end{abstract}

\maketitle

\clearpage

%
%
\section{Introduction}

The algebraic approach to relativistic quantum field theory (RQFT) \cite{Haag1},\cite{Haag2} is obtained by characterizing the algebras ${\cal{A}}({\cal{O}})$ of local observables we can define on a bounded region ${\cal{O}}$ of space-time. In this algebraic approach any model of a RQFT is defined by a net ${\cal{O}} \rightarrow {\cal{A}}({\cal{O}})$ of von Neumann algebras acting on a Hilbert space 
${\cal{H}}$ \cite{Haag3}, \cite{Haag4}, \cite{Haag5}. A very basic result of this algebraic approach is that the algebra ${\cal{A}}({\cal{O}})$ for ${\cal{O}}$ any bounded open set of space-time and for any RQFT are hyperfinite type $III_1$ factors in Murray-von Neumann classification \cite{vN} and \cite{Murray} \footnote{For a set of excellent technical lectures on this and related topics see \cite{Witten1},\cite{Witten2}. For a less technical approach see\cite{Casini}}. Since these sort of factors are all isomorphic it means that, from the point of view of local observables, RQFT is {\it unique}. This {\it universality} has several important physical consequences. The simplest is that any state can be obtained, with arbitrary precision, by means of local operations \cite{RS}. This can rephrased in more familiar terms saying that, in any RQFT, we cannot associate with a local bounded region of space-time neither the Hilbert space of quantum micro states describing that region nor even a density matrix. This fact is related to the {\it universal} ultra violet divergences of the quantum entanglement \cite{sorkin} between a bounded region ${\cal{O}}$ and its causal complement.

When we work with RQFT on a curved space-time with {\it horizons}, either black hole horizons or cosmological horizons, the corresponding algebras of local observables for the observer in the black hole exterior or for the observer inside the cosmological horizon are indeed type $III$ factors. This roughly means that within the context of RQFT the observer outside the black hole cannot assign to the black hole interior any form of Hilbert space of micro states not even a density matrix or equivalently, for cosmological horizons, that the observer inside the cosmological horizon cannot describe her physics in terms of quantum micro states or even a density matrix. This RQFT result contrast with the familiar Bekenstein-Hawking expressions for the black hole entropy \cite{Bekenstein}, \cite{Hawking} or with the Gibbons-Hawking expression \cite{GH} for the entropy of de Sitter space-time.

The deep meaning of this apparently severe limitation of RQFT is connected to another basic result of type $III_1$ factors known as Tomita-Takesaki theory \cite{Tomita1},\cite{Tomita2},\cite{Tomita3},\cite{Tomita4}. These factors come equipped with a unique and well defined notion of {\it time} translation implemented as an automorphism of the algebra of local operators ${\cal{A}}({\cal{O}})$ \footnote{See \cite{Roveli} and \cite{Connes}.} . The corresponding generator, that we will denote $K$ is known as {\it modular Hamiltonian}. This time evolution is however an {\it outer} automorphism which in more plain words simply means that the modular Hamiltonian is not part of the algebra of local observables. Again the absence  of a local representation of the natural generator of time translations associated with the algebra ${\cal{A}}({\cal{O}})$ is reflecting the divergent entanglement between
${\cal{O}}$ and its causal complement.

Our basic intuition about quantum gravity, grounded on the finite entropies we associate with black holes and cosmological horizons, seems to indicate that the key role of quantum gravity effects would be to regulate somehow the infinite entanglement implicit in type $III$ factors transforming the corresponding algebra into a type $I$ or type $II$ factor i.e. an algebra with a well defined Hilbert space of quantum microstates or at least a well defined density matrix. How to do that without violating the basic axioms of RQFT on which the type $III$ nature of the algebra ${\cal{A}}({\cal{O}})$ is based ?

Recently two very interesting proposals to deal with this issue have been suggested. On one side Leutheusser and Liu \cite{LL1},\cite{LL2} have considered the possibility to define new emergent times, different from the natural modular time, using Borcher's half modular inclusion \cite{Borchers}, \cite{Borchers2}. The crucial aspect of this emergent times is that it could allow us to define time flows that go beyond the horizon. More precisely, and relative to the generator defining evolution in the emergent time, we can have real causal exchanges between the left and right exterior regions of the eternal AdS black hole. In more simple words what this means is that states defined by a local physical perturbation in ${\cal{A}}({\cal{O}})$ can be, after evolving in the new emergent time, quantum mechanically distinguished using the commutant algebra i.e. the algebra of local observables on the causal complement of ${\cal{O}}$.

The other proposal by Witten \cite{witten} is based on using the notion of crossed products ( for a good review see \cite{Phillips} ) \footnote{This approach has been recently extended to de Sitter space-time in \cite{witten2}}. Heuristically the idea of crossed product is to extend the algebra ${\cal{A}}({\cal{O}})$ in such a way that the time translation generator is now associated with an {\it inner} automorphism of the extended algebra. Once this extension from {\it outer time} into {\it inner time} is done the extended algebra becomes a type $II$ factor. In a nutshell, this crossed product transformation of a type $III$ factor into a more familiar type $II$ factor, where we can use the notion of von Neumann entropy, is done using, as elements of the extended algebra, continuous and bounded functions ("paths") from time into the algebra ${\cal{A}}({\cal{O}})$ i.e. elements in $L^2(R,{\cal{A}}({\cal{O}}))$ and, using as Hilbert space, the space of paths in $L^2(R,{\cal{H}})$. Very likely there exist a deep connection between the emergent time proposal \cite{LL1},\cite{LL2} and the crossed product approach \cite{witten},\cite{witten2} in the form of a natural relation between the emergent time extension of the domain of the algebra and a crossed product extension where we use paths in the emergent time.

Coming back to the basic intuition about the role of gravity as regulator we should try to figure out how quantum gravity effects can be encoded in the crossed product  structure. In other words how gravity can resolve the elements contributing to the states in $L^2(R,{\cal{H}})$ representing the Hilbert space for the extended algebra.

Heuristically, and although the modular generator $K$ leaves invariant the cyclic ground state used to represent ${\cal{A}}({\cal{O}})$, this is not the case for the states defining a non trivial path in $L^2(R,{\cal{H}})$. This simple fact allows us to evaluate the quantum variance $\Delta(K^2)$ for the different elements defining the state in $L^2(R,{\cal{H}})$. 

After coupling to gravity this quantum variance will {\it gravitate}  generating a {\it gravitational potential}. In this note we will use this general approach to define the spectrum of scalar curvature fluctuations in Inflationary Cosmology.

In order to frame Cosmology in terms of a crossed product we will describe the inflationary epoch using a state in the natural crossed product Hilbert space $L^2(R^+,{\cal{H}}_{dS})$ with ${\cal{H}}_{dS}$ the GNS Hilbert space representation of the algebra associated with the static patch and with $R^+$ representing the conformal time during the inflationary epoch. In order to identify this state we will start with QFT for a free scalar defined on the planar patch of de Sitter. Relative to the conformal time we will interpret the time dependent algebra of creation annihilation operators defined by the Bogolyubov transform as well as the conformal time dependent vacuum states for the different modes, as elements in $L^2(R^+,{\cal{H}}_{dS})$.  For these states we evaluate the quantum variance for the generator of translations in the conformal time and we work out the gravitational potential induced by this quantum variance. This defines the scalar curvature fluctuations.

This program for the computation of scalar curvature fluctuations can be developed without introducing any classical quasi de Sitter model for the inflaton potential \cite{GJ1}\cite{GJ2} and leads to very concrete numerical predictions both for the spectral index 
\begin{equation}
(1-n_s)= 0.0328
\end{equation}
 as well as for the slow roll parameter 
 \begin{equation}
 \epsilon \sim 0.0027
 \end{equation}

Thus this approach hopefully increases substantially the predictive power of inflationary Cosmology.

As an aside comment let us just observe that the approach based on evaluating the gravitational potential induced by the quantum variance of the operator generating time translations share close similarities with the recent proposal for the computation of entropy fluctuations in causal diamonds \cite{Verlinde1},\cite{Verlinde2} \cite{banks}.

\section{Horizons}
In order to setup the formal frame in which we will be working it could be convenient to start with a very general and basic question, namely : How to define a subsystem in quantum physics ? The natural answer if we know the total Hilbert space ${\cal{H}}$ is to {\it split} this space as ${\cal{H}}_A \otimes {\cal{H}}_B$ with ${\cal{H}}_A$ representing the Hilbert space of the subsystem. This decomposition of the Hilbert space can be easily done if we are dealing with finite number of degrees of freedom and we decide to define the subsystem using a concrete subset of the full set of canonical variables. The unique irrep of this subset of canonical variables defines automatically ${\cal{H}}_A$ and the desired decomposition. This decomposition is unique due to the well known theorem by von Neumann that essentially says that the quantum mechanics of a finite set of degrees of freedom is unique i.e. only one irrep up to unitary equivalence. To split the Hilbert space is very useful to define the degree of entanglement between the subsystem and its environment. Indeed, if we have a bounded Hamiltonian describing the dynamics of the full system and we define the full ground state $|\Phi_0\rangle$ then a good measurement of the entanglement is given by the von Neumann entropy $S=-Tr(\rho_A \ln \rho_A)$ for $\rho_A= Tr_{{\cal{H}}_A} |\Phi_0\rangle \langle \Phi_0|$. Can we extend this picture when we move into quantum field theory or when we work with statistical systems in the thermodynamic limit ? The answer to this query is negative. 

In order to gain some intuition on what can be the problem let us consider the following simple exercise \footnote{The following example is based on a paradox initially pointed out by Fermi \cite{Fermi}. For a modern description using type III factors see \cite{Fermi2},\cite{Fermi3} and \cite{Fermi4}.}. Imagine your localized subsystem and consider a local perturbation taking place outside. Now you wonder if, for a given Hamiltonian describing the time evolution, the perturbation outside can be detected, after a certain time, measuring properties of the subsystem. Let us take as initial state one where the subsystem is in its ground state and denote this state $|\Phi\rangle$. The {\it question}: Is this subsystem perturbed after a time $t$ ? can be associated, in quantum mechanics, to a projection $P_A$ whose eigenvalue, corresponding to answer Yes or Not, can be discovered performing measurements on the corresponding subsystem. Now if the dynamics defined by the Hamiltonian, where any form of interaction between the subsystem and the environment can be implemented, is {\it bounded}, then you can prove that $f(t)$ defined by $\langle \phi|P_A e^{iHt}|\Phi\rangle$ for any $|\phi\rangle$ can be analytically continued in time to the upper half plane. This simple fact has a dramatic consequence, namely that if $f(t)$ is zero for some small interval of time then it is identically zero. If we are thinking in an initial state with the perturbation localized far away from the subsystem we expect, on the basis of {\it locality}, that some finite time is needed to propagate this perturbation to the subsystem and therefore we should assume,if the Hamiltonian does not contain superluminal effects, that $f(t)$ is zero for some finite interval. So the former analyticity argument implies that $f(t)$ should be identically zero. 

We are mentioning this paradox to provide some preliminary intuition on the necessity to define subsystems using the notion of Murray and von Neumann factors \cite{Murray}. Indeed, the former paradox has its origin in the use of the minimal projection $P_A$. In other words what this example is indicating is that a definition of subsystem in local quantum field theory requires that the algebra of operators associated with local measurements on the subsystem should be an algebra without minimal projections. This is the first big surprise. Indeed if the algebra of local observables characterizing the subsystem has not minimal projections we cannot have a notion of pure quantum state to be associated with the subsystem and consequently no notion of the sub Hilbert space ${\cal{H}}_A$. In summary no split property of the full Hilbert space.

The algebraic approach of Murray and von Neumann is to define subsystem in terms of the properties of the algebra of local observables, let us call it ${\cal{A}}$. This algebra should be a von Neumann algebra, which means that is an algebra of bounded operators acting on the full Hilbert space closed under the adjoint operation and complete in the state dependent weak topology \footnote{In this topology you define the limit of $a_n$ as $a$ if for any state in the Hilbert space $lim\langle \psi|a_n|\psi\rangle = \langle \psi|a|\psi\rangle$. The condition of being closed under this state topology is equivalent to the most familiar characterization of von Neumann algebras as those that  are equal to its double commutant.} . In addition you require that this algebra is a {\it factor}, which means that it has a trivial center. Under these conditions you can prove that ${\cal{B}}({\cal{H}}) = {\cal{A}} \otimes {\cal{A}}'$ with ${\cal{A}}'$ the commutant i.e. the set of operators commuting with all elements in ${\cal{A}}$. Note the underlying logic. Instead of looking for a split of the Hilbert space we look for a split of ${\cal{B}}({\cal{H}})$. In this sense the classification  of types of {\it quantum subsystems} becomes the classification of factors. This classification ( recall the former discussion) can be obtained by analyzing the type of projections. Type $I$ are factors containing minimal projections. In this case they are associated with a Hilbert space of states and they are the factors that describe subsystems with finite number of degrees of freedom. Now type $II$ and type $III$ are more exotic. In case of type $II$ there exist finite projections but not minimal projections. In this case we can assign to the subsystem a density matrix but not pure quantum microstates. In the case $III$ we don't have neither minimal not finite projections. This rather exotic case is however the one describing localized subsystems in local quantum field theory. In order to motivate this result we presented the argument about the analyticity of $f(t)$ above.

Given  a $C^*$ algebra ${\cal{A}}$ that we physically identify as the algebra of observables describing a localized subsystem the associated quantum physics is determined once we have a {\it representation} of this algebra in terms of bounded operators on a Hilbert space. This Hilbert space is the {\it quantum arena} defining the space of states. If ${\cal{A}}$ is a type $I$ factor there exist a unique irrep of ${\cal{A}}$ as the space of bounded operators on some Hilbert space that is the Hilbert space ${\cal{H}}_A$ of states of the subsystem. Although we can now think of ${\cal{H}}_A$ as a subspace of some bigger Hilbert space representing the subsystem and the rest the irrep of ${\cal{A}}$ is independent of how this subsystem is part of the bigger system. However when ${\cal{A}}$ is a type $III$ factor the corresponding representation, that is known as GNS representation and that we will review in a moment, defines simultaneously the representation of ${\cal{A}}$ and of its commutant ${\cal{A}}'$. Physically what this means is that the GNS representation of the algebra already contains full information on the entanglement of the subsystem. This algebraic implementation of the entanglement is the key property of type $III$ factors. Physically we could say that the algebra {\it knows a priori} about its {\it environment} and the corresponding entanglement \footnote{That the entanglement is encoded in the algebra and not depending on the state is a basic property of type $III$ factors that we will use many times. Intuitively this fact explains the universality of UV divergences of the entanglement in local quantum field theory.}.

In addition for type $III$ factors Tomita-Takesaki theory \cite{Tomita1},\cite{Tomita2},\cite{Tomita3}, \cite{Tomita4} provides, for each GNS representation a modular automorphism of the algebra that we can formally interpret as defining a sort of {\it time evolution}. The generator of this time evolution is the well known {\it state dependent} modular Hamiltonian. If we think ${\cal{A}}$ as the algebra of local observables associated with a region $E$ of space time what Tomita Takesaki (TT) theory is indirectly providing is a global time defined on $E$ and a time flow that acts as automorphism of ${\cal{A}}$. This automorphism is also acting on ${\cal{A}}'$ and in that sense this modular time evolution prevents the existence of any causal correlation in time between ${\cal{A}}$ and ${\cal{A}}'$. In physics terms what we get from the GNS representation and TT-theory is a subsystem represented by ${\cal{A}}$ and an environment represented by ${\cal{A}}'$ causally disconnected in modular time flow although infinitely entangled. 

Heuristically we could think that the space time region $E$ associated in this case to the algebra ${\cal{A}}$ is bounded by some {\it horizon} that cannot be crossed flowing in modular time. This heuristic interpretation of horizon as the boundary of the space time region with an associated algebra of local operators of type $III$ nicely works, as recently described by \cite{LL1}, \cite{LL2}  for the exterior regions of the eternal AdS black hole but also for the static patches of de Sitter space \cite{witten},\cite{witten2}. 

\subsection{The GNS representation}
Let us start considering an abstract $C^*$ algebra ${\cal{A}}$ and let us define a linear functional $f:{\cal{A}} \rightarrow R$. Associated with this functional $f$ we define the ideal ${\cal{I}}_{{\cal{A}},f}$ as the set of elements  $y$ in ${\cal{A}}$ such that:
\begin{equation}
f(y^{+}y) =0
\end{equation}
Let us now define the space of equivalence classes $\frac{{\cal{A}}}{{\cal{I}}_{{\cal{A}},f}}$ where two elements in ${\cal{A}}$ are in the same equivalence class if they differ by an element in ${\cal{I}}_{{\cal{A}},f}$. We will denote as $[a]$ the equivalence class of the element $a$. Let us now define a Hilbert space ${\cal{H}}^{GNS}_{{\cal{A}}}$ where we associate to each equivalence class a vector $|[a]\rangle$ with the scalar product defined as
\begin{equation}
\langle [a]|[b]\rangle = f(a^{+}b)
\end{equation}
In this sense ${\cal{H}}^{GNS}_{{\cal{A}}} \sim \frac{{\cal{A}}}{{\cal{I}}_{{\cal{A}},f}}$.
Note that $f$ defined on $\frac{{\cal{A}}}{{\cal{I}}_{{\cal{A}},f}}$ satisfies the conditions defining a {\it state} i.e. a linear form such that $f(a^{+}a) >0$ and $f(1)=1$. The GNS representation of ${\cal{A}}$ in the space of bounded operators of ${\cal{H}}^{GNS}_{{\cal{A}}}$ is defined by
\begin{equation}
\pi(a) |[b]\rangle = |[ab]\rangle
\end{equation}
Let us denote $|\Phi_0\rangle$ the element in the Hilbert space ${\cal{H}}^{GNS}_{{\cal{A}}}$ corresponding to the equivalence class of the identity i.e. $|[1]\rangle$. 

Once we have defined ${\cal{H}}^{GNS}_{{\cal{A}}}$ we can define the commutant ${\cal{A}}'$ as the algebra of bounded operators in ${\cal{H}}^{GNS}_{{\cal{A}}}$ commuting with all the elements in ${\cal{A}}$. 

It is now easy to check that the state  $|\Phi_0\rangle$ is cyclic i.e. ${\cal{A}}|\Phi_0\rangle$ is dense in ${\cal{H}}^{GNS}_{{\cal{A}}}$ and separating i.e. $a|\Phi_0\rangle=0$ implies $a=0$ for both ${\cal{A}}$ and its commutant ${\cal{A}}'$.

The crucial aspect of the GNS representation is that the state associated with the identity is not only cyclic as we are use to think the vacuum in Fock space but also separating. Is this property, absent for the standard Fock vacuum, what makes this state to be also cyclic with respect to the commutant of the algebra.  

Intuitively the fact that the GNS Hilbert space can be interpreted as the completion of ${\cal{A}}|\Phi_0\rangle$ as well as that of ${\cal{A}}'|\Phi_0\rangle$ seems to imply the existence of some sort of symmetry transformation that, leaves the state $|\Phi_0\rangle$ invariant, and  maps elements in ${\cal{A}}$ into elements in the commutant ${\cal{A}}'$. This has important physical consequences that become clear once we introduce modular Tomita's operators.

\subsection{Modular Dynamics}
Given a GNS representation of the algebra ${\cal{A}}$ the Tomita operator is defined by the relation
\begin{equation}\label{tomi1}
S_{f,{\cal{A}}} a |\Phi_0\rangle = a^{+} |\Phi_0\rangle
\end{equation}
for every element $a$ in ${\cal{A}}$. As an operator acting on ${\cal{H}}^{GNS}$ this is an unbounded operator with dense domain. We can equally define the analog for the commutant ${\cal{A}}'$ as
\begin{equation}\label{tomi2}
S^{+}_{f,{\cal{A}}} b |\Phi_0\rangle = b^{+} |\Phi_0\rangle
\end{equation}
for every $b$ in ${\cal{A}}'$. The modular operator is defined as
\begin{equation}
\Delta_{f,{\cal{A}}} = S^{+}_{f,{\cal{A}}} S_{f,{\cal{A}}}
\end{equation}
which implies that
\begin{equation}\label{Tomi1}
S_{f,{\cal{A}}}= J_{{\cal{A}}} \Delta^{1/2}
\end{equation}
and
\begin{equation}\label{Tomi2}
S^{+}_{f,{\cal{A}}}= J_{{\cal{A}}} \Delta^{-1/2}
\end{equation}
with $J_{{\cal{A}}}$ satisfying $J^2=1, J{\cal{A}}J={\cal{A}}'$ and $J\Delta^{1/2} J = \Delta^{-1/2}$.

The operator $J$ represents the {\it symmetry} transformation we were looking for, namely it maps ${\cal{A}}$ into ${\cal{A}}'$. 

For arbitrary real values of $t$ we can define an hermitian operator acting on ${\cal{H}}^{GNS}$ as $\Delta^{it}$. The adjoint action
\begin{equation}
\Delta^{it} a \Delta^{{-it}}
\end{equation}
defines an {\it outer automorphism} of the algebra ${\cal{A}}$.This is an outer automorphism if the corresponding generator $K_{{\cal{A}}} = log \Delta_{{\cal{A}}}$ is not part of the algebra ${\cal{A}}$. This automorphism as well as the operator $K_{{\cal{A}}}$ depends on the particular state $|\Phi_0\rangle$ used in the corresponding GNS construction. It can be proved that changes in the choice of the cyclic state correspond to inner automorphisms \cite{Tomita2}. 

In what follows we will refer to the dimensionless parameter $t$ as {\it modular time} and to the generator $K_{{\cal{A}}}$ as {\it modular Hamiltonian} \footnote{On the physical meaning of the modular time see \cite{Roveli} and \cite{Connes}}. Note that with respect to this modular time the operator $J$ introduced above plays the role of {\it time reversal}. This gives us the clue on the hidden thermal properties of cyclic and separating states. 

Indeed from (\ref{Tomi1}) and (\ref{Tomi2}) we derive $\Delta^{ 1/2}=J S$ and $\Delta^{-1/2} = JS^{+}$. Thus the correlator ( in formal ket bra notation ) $\langle\Phi_0|b \Delta^{it} a |\Phi_0\rangle$ for generic elements $a$ and $b$ can be analytically extended to $t-i$ by replacing $\Delta^{it}$ by $\Delta^{1/2}\Delta^{it} \Delta^{1/2}$. Using now the definition of the Tomita operators (\ref{tomi1}) and (\ref{tomi2}) we get
\begin{equation}\label{KMS}
\langle\Phi_0|b \Delta^{i(t-i)} a |\Phi_0\rangle = \langle Jb \Phi_0|J\Delta^{it} a^{+} \Phi_0\rangle = \langle \Phi_0|a\Delta^{{-it}}b|\Phi_0\rangle
\end{equation}
where the role of $J$ as time reversal becomes manifest in the last step. This identity is the definition of KMS states for von Neumann algebras.

Recall that in quantum mechanics we say that a given state is a KMS thermal state at inverse temperature $\beta$ relative to the time evolution defined by a given Hamiltonian $H$ if the expectation values $f(t) = \langle B A(t) \rangle$ for $A(t) = e^{iHt}Ae^{-iHt}$ and $B$ generic is analytic in the imaginary interval $0< Im t < \beta$ and satisfies (\ref{KMS}) i.e $\langle A(t) B \rangle = \langle B A(t+i\beta) \rangle$ . This is equivalent to say \cite{Haag} that the KMS state is defined by the canonical thermal density matrix $\rho = e^{-\beta H}$ appropriately normalized \footnote{As pointed out in \cite{Witten1} the KMS property that relates different operator orderings through analytic continuation is crucial in studies of quantum chaos \cite{Malda2} and complexity growth.}. 

Naively we could try to make this correspondence explicit by identifying the GNS value $f([a])$ with $tr([a]\rho)$ for some $f$-dependent hamiltonian $\hat H$ and inverse temperature $\beta$. This is not working when we deal with type $III$ factors for which we cannot define a trace. This is in particular the case when we work in the thermodynamic infinite volume limit or in local QFT. However the relation to KMS states can be defined using only the analyticity properties. This leads to the operator map 
\begin{equation}
\Delta_{{\cal{A}},f} = e^{-\beta \hat H}
\end{equation}
or more explicitly $K= -\beta \hat H$. The natural question at this point is: What is the physical meaning of $\hat H$?

The first step into a physical understanding of the meaning of the modular generator $K_{{\cal{A}}}$ was done by Bisognano and Wichmann \cite{BW} using the algebra of local operators associated with a Rindler wedge. This is an specially illustrative example for several reasons. In the Rindler frame we can divide Minkowski into two wedges to be denoted $L$ and $R$. Now we can check that the GNS representation of the local algebra associated with, let us say, the $L$ wedge, has as commutant the local algebra of operators on the $R$ wedge. That the modular generator $K$ becomes the boost generator on the $L$ wedge with the modular time the {\it rapidity time} parametrizing the trajectories. Moreover the transformation $J$ defined above becomes geometrically the antipodal map between the two wedges and finally the cyclic and separating state defines the analog of the Unruh vacuum. In other words the KMS thermality of the GNS state associated to the algebra of local operators defined on the Rindler wedge becomes the thermality discovered by Unruh \cite{Unruh} for an accelerated observer.

\subsection{Purification}
The natural relation between KMS thermal states and the cyclic and separating states of the GNS representation of a von Neumann algebra ${\cal{A}}$ could motivate us to look for a {\it purification} of the operator $\rho_{\cal{A}} = e^{K_{\cal{A}}}$ defined in terms of the modular Hamiltonian $K_{\cal{A}}$. 

We will generically define a {\it Fock purification} of the algebra ${\cal{A}}$ in terms of a couple of isomorphic commuting Heisenberg algebras: ${\cal{A}}_a$ generated by $a_{k}, a^{\dagger}_{k}$ and ${\cal{A}}_b$ generated by $b_{k}, b^{\dagger}_{k}$ and an anti linear operator $J$ such that $Ja_kJ=b_k$. Associated to this couple of Heisenberg algebras we define a Fock space representation ${\cal{F}}= {\cal{F}}_a \otimes {\cal{F}}_b$. We will say that a given pure state $|\beta\rangle \in {\cal{F}}$ is a Fock purification of ${\cal{A}}$ if 
\begin{equation}
{\cal{A}} = \frac{{\cal{A}}_a}{I_{|\beta\rangle}}
\end{equation}
and
\begin{equation}
{\cal{A}}' = \frac{{\cal{A}}_b}{I_{|\beta\rangle}}
\end{equation}
for the ideal $I_{\beta}$ defined as the set of elements $x\in {\cal{A}}_a \otimes {\cal{A}}_b$ such that $\langle \beta|x^{\dagger}x|\beta\rangle =0$. The GNS Hilbert space is defined by the set of states $|[a]\rangle$ with $a\in {\cal{A}}_a$ for $[a]$ the equivalence class defined relative to $I_{|\beta\rangle}$ with scalar product $\langle [a]|[a']\rangle = \langle\beta|a^{\dagger}a'|\beta \rangle$. The cyclic and separating state $|\Phi_0\rangle$ of this GNS representation is associated with the equivalence class in ${\cal{F}}$ defined by the state $|\beta\rangle$ i.e. the set of states of the form $(1+x)|\beta\rangle$ for $x\in I_{|\beta\rangle}$. Now the Tomita operator defined as
$S[a]|\Phi_0\rangle = [a^{+}]|\Phi_0\rangle$ and $S^+[b]|\Phi_0\rangle = [b^+]|\Phi_0\rangle$  can be directly derived from the action of $a \in {\cal{A}}_a$ and $b\in {\cal{A}}_b$ on $|\beta\rangle$. 

What we have described is just a version of the TFD formalism where the two Heisenberg algebras represent the two copies of the same system. In order to get the explicit expression of the Tomita operators we need to start with a concrete state $|\beta\rangle$. The TFD state for an inverse temperature $\beta$ could be formally defined as
\begin{equation}
|\beta\rangle = \frac{1}{\sqrt{N}}\sum_n e^{-\beta E(n)/2} |n\rangle_a |n\rangle_b
\end{equation}
where generically the state $|n\rangle_{a,b}$ represents a basis of energy eigenstates in ${\cal{F}}_{a,b}$ with energy $E(n)$. We can think these states described by distributions $n(k)$ with $n(k)$ the number of $a^{\dagger}_k (b^{\dagger}_k)$ acting on the Fock vacuum and $E(n) = \int \omega(k)n(k)$ for $\omega(k)$ the "energy" associated (created) with the modes $a^{\dagger}_k / b^{\dagger}_k$. Obviously this energetics will be dependent on the particular physics model we use to introduce the two Heisenberg algebras. 

In this representation is easy to check
\begin{equation}\label{Bogo1}
a_k^{+} |\beta\rangle = e^{\beta \omega(k)/2}b_k|\beta\rangle
\end{equation}
and
\begin{equation}\label{Bogo2}
b_k^{+}|\beta\rangle = e^{\beta \omega(k)/2}a_k|\beta\rangle
\end{equation}
Comparing now with (\ref{tomi1}) and (\ref{tomi2}) we easily get the representation of the modular generator $K_{\cal{A}}$ as
\begin{equation}\label{puri}
K_{\cal{A}} = \beta (H_b-H_a)
\end{equation}
where we define the formal Hamiltonians $H_{a,b}$ as formally measuring the total energies of the basis states used in the definition of the state $|\beta\rangle$. Note that these energies can be divergent in the thermodynamic limit. However the modular operator
is well defined \footnote{The main point of this purification exercise is to observe how in this context the Tomita operators as defined by (\ref{Bogo1}) and (\ref{Bogo2}) reflect the entanglement implicit in the state $|\beta\rangle$. In other words is the entanglement of this state defining the purification what generates the modular time flow.} \footnote{Note that for a given Fock space purification the GNS Hilbert space is a subspace that in the language of quantum error correcting codes could be interpreted as a code subspace.}. 

\subsection{Space-Time interpretation}
The GNS construction is specially suited to associate a space-time interpretation to a given von Neumann algebra ${\cal{A}}$. First of all the modular automorphism $\Delta^{it}$ formally defines a time flow for the modular time $t$. Since $\Delta^{it}{\cal{A}} \Delta^{-it} = {\cal{A}}$ and the same for the commutant ${\cal{A}}'$ we can associate the Heisenberg's trajectories $a(t) = \Delta^{it} a \Delta^{-it}$ for $a\in{\cal{A}}$ with paths in ${\cal{A}}$. Thus if ${\cal{A}}$ admits the interpretation of the algebra ${\cal{A}}(E)$ of local operators with support in a bounded region of space time $E$ the GNS construction defines on $E$ a global "time" i.e. a global Killing, as well as the set of trajectories $a(t)$ defined by the modular flow. Moreover the GNS construction provides the antilinear map $J$ and therefore we can induce an associated space time transformation $\hat J$ in such a way that the commutant of ${\cal{A}}(E)$ can be represented as ${\cal{A}}(\hat J (E))$. The modular time flow on the region $\hat J(E)$ is the time reversal of the modular time flow defined on region $E$.

With this information at hand we can try to look for a formal space-time geometry where we can define a global Killing time on regions $E$ and $\hat J(E)$ related by time reversal. We can figure out some examples of this situation using Penrose diagrams and the basic intuition provided by Bisognano Witchman construction \cite{BW}. Two interesting examples are the ones defined by the Penrose diagram of the eternal AdS black hole and the one associated with eternal de Sitter.

In the case of the eternal AdS black hole the Penrose diagram is a square with two diagonals representing the horizon. The upper horizontal line represents the singularity at $r=0$ while the two vertical boundary lines represent the asymptotic $r=\infty$ regions.  The two diagonal lines cross at the bifurcation point in the center. In holographic language \cite{Malda1} the two lateral triangular regions have a holographic dual CFT defined on the corresponding  boundary. We will denote these regions $L$ and $R$- exterior and the corresponding CFT's $CFT_{R,L}$. In these regions we can define a global time that is the time of the CFT's. Both regions are related by {\it the antipodal map}. The upper and low triangular regions represent respectively the future and past interiors.  In the left and right exterior regions we can define two algebras of local operators ${\cal{A}}_L$ and ${\cal{A}}_R$ that we will identify with the algebras of single trace operators \cite{LL2} in the corresponding CFT's. 

The GNS representation of ${\cal{A}}_R$ identifies the global time in the $R$- exterior with the modular time, the commutant with ${\cal{A}}_L$ and $\hat J$ with the antipodal map between the $R$ and $L$ exterior regions. In this case the parts of the Penrose diagram not described by the GNS representation of 
${\cal{A}}_R$ correspond to the upper and lower interiors of the black hole. In this case the holographic description of these algebras as the algebras of single trace operators of the boundary CFT's is natural since the vertical boundaries of the Penrose diagram represent the asymptotic infinity at $r=\infty$. 

In the case of pure de Sitter the Penrose diagram is very similar with two important differences. The vertical boundary lines correspond now to $r=0$ and the boundary horizontal line to $r=\infty$ with the two diagonals representing the horizon. The former $L$ and $R$ regions define the two static patches related by the antipodal map. How to define in this case the holographic map is an open and interesting question that we will not touch in this section. 

What we observe with these examples is that the GNS construction naturally covers a region of the Penrose diagram leaving outside the black hole interior in the AdS case or the exterior of the cosmological horizon in the de Sitter case. Moreover the two regions described by the GNS construction related by the modular anti linear map $\hat J$ are, from the point of view of the modular dynamics, {\it causally disconnected}.

The {\it modular } causal independence simply means that the time evolution defined by the modular dynamics will never create any {\it physical energy  exchange} between these two regions. This fact can be rephrased as {\it the indistinguishability using observations associated with operators in ${\cal{A}}_R$ of the cyclic state $|\Phi_0\rangle$ and the state in the GNS Hilbert space $|b\rangle$ associated with any unitary operator $b$ in the commutant i.e. ${\cal{A}}_L$.}

 In a formal ket-bra notation this statement reduces to the equality
\begin{equation}
\langle \Phi_0|O|\Phi_0\rangle = \langle b|O|b\rangle
\end{equation}
for any $O \in {\cal{A}}_R$. Moreover this is also the case if we evolve $b$ in the modular time $t$ using $\Delta^{it}$ i.e.
\begin{equation}
\langle \Phi_0|O|\Phi_0\rangle = \langle b(t)|O|b(t)\rangle
\end{equation}
for any $t$.

The former result can be made more precise using Araki's notion of relative entropy (see \cite{Witten1} ) ${\cal{S}}(|\Phi_0\rangle || |b(t)\rangle)$ relative to the algebra ${\cal{A}}_R$ \cite{araki}. This quantity is given by:
 \begin{equation}
 {\cal{S}}(|\Phi_0\rangle || |b(t)\rangle)= - \langle \Phi_0| \log (\Delta(\Phi_0|b(t)) |\Phi_0\rangle
 \end{equation}
 with the relative modular operator $\Delta(\Phi_0|b(t))$ defined as
 \begin{equation}
 \Delta(\Phi_0|b(t)) = S^{\dagger}_{\Phi_0|b(t)}S_{\Phi_0|b(t)}
 \end{equation}
 and with the Tomita operator $S_{\Phi_0|b(t)}$ defined by
 \begin{equation}
 S_{\Phi_0|b(t)} a |\Phi_0\rangle = a^{\dagger}|b(t)\rangle
 \end{equation}
 for $a$ any element in ${\cal{A}}_R$ and $S^{\dagger}_{\Phi_0|b(t)}$ equally defined  but for $a$ any element in the commutant of ${\cal{A}}_R$. Note that the state $|b(t)\rangle$ is not cyclic. Now we can easily prove that
 \begin{equation}\label{two}
 {\cal{S}}(|\Phi_0\rangle || |b(t)\rangle) =0
 \end{equation}
 for any time $t$ provided $b(t)$ is in the commutant of ${\cal{A}}_R$ as it is the case if the time flow is defined by the modular automorphism. Thus the relative entropy ${\cal{S}}(|\Phi_0\rangle || |b(t)\rangle)$ measures the distinguishability between the states $|\Phi_0\rangle$ and the state $b(t)\rangle$ using local observations associated with elements in ${\cal{A}}_R$. 
 
 The obvious causality meaning is that any physical perturbation created in the $L$ region cannot be detected using local operators in the $R$ region. In order to stress the physical meaning of the perturbation we have used unitary operators. \footnote{If we consider instead of $b$ a non unitary operator $a$ we could distinguish the states. However in this case the state $|a\rangle$ is not representing any physical energetic perturbation. This clarifies the causality meaning of the Reeh-Schlieder theorem \cite{RS}. Indeed we can always find an operator $a$ in ${\cal{A}}_R$ such that the GNS state $|a\rangle$ approach as much as you wish the state $|b\rangle$ however the difference in this case between $\langle \Phi_0|O|\Phi_0\rangle$ and   $\langle a(t)|O|a(t)\rangle$ is not reflecting any causal interplay between both regions} \footnote{The causality problem described above is similar to the one appearing in Fermi's problem. For the case of the eternal AdS black hole this problem was the main target of \cite{LL1}. As we will briefly discuss the suggested solution in \cite{LL1} and \cite{LL2} was to use an {\it emergent new time} and to show that by flowing in this emergent time we can distinguish, after some critical time, the state associated with a unitary perturbation created in the left region $L$,  from the cyclic ground state, using measurements associated with the local algebra on the right side $R$. In other words, it was suggested that there exist new emergent times that can allow us to inspect the black hole interior.}
 
\subsection{Modular generator and "energies"}
In the previous discussion, think in the eternal AdS black hole case, we have a modular time flow with generator $K$. Geometrically $K$ induces opposite time translations on the left and right side of the Penrose diagram. In addition we could think on the independent generators of time translations on the left and right side. These generators are the ones we can formally denote $H_R$ and $H_L$ on the basis of our discussion on the TFD state. They are associated to a Killing that, for the $H_R$ case, is $\partial_t$ on the right side and zero on the left side. Such operator will act on ${\cal{A}}$ if we associate ${\cal{A}}$ to the right exterior and will not act on ${\cal{A}}'$ that, in this case we associate with the left exterior. This form of action is indeed ill defined when we work with type $III$ factors. The physics reason is easy to understand. Indeed since in this case we have infinite entanglement between both regions we cannot define, using time evolution, independent actions on the L and R exteriors, as the ones we represent by $H_R$ and $H_L$. Technically what this means is that these operators are not well defined in the sense that $||H_{R,L}|\psi\rangle||^2$ for any $|\psi\rangle$ in $H^{GNS}$ is infinity. This is not the case for the modular generator $K$ that acts simultaneously ( although not in a factorized way) on both ${\cal{A}}$ and ${\cal{A}}'$. 

An important question is if we can {\it regulate} the definition of $H_{R,L}$. Since the divergence of $||H_{R,L}|\psi\rangle||^2$ has the same origin as the divergence of the entanglement we can try to regulate this divergence including some UV cutoff. The natural UV cutoff will be the Planck mass once we couple the system to gravity. So we can expect that the expectation values of $H_{R,L}$ diverge as a power of $M_P$. What this actually means is that these operators are ill defined once we decouple gravity i.e. once we send $M_P=\infty$.

At this point the reader should be aware of the following potential puzzle. Imagine we invent a UV regulator such that we can give perfect sense to $H_{R,L}$. However due to the infinite entanglement between ${\cal{A}}$ and ${\cal{A}}'$ the only natural possibility is that these regulated operators are commuting with the algebra and with its commutant. In other words they should represent central elements characterizing classical geometrical properties of the space time background. However adding central elements perverts the nature of the algebras as factors. Thus the logic, underlying how to regulate the UV infinite entanglement, should follow a basic recipe, namely {\it to regulate preserving the nature of the local algebra as a factor.} This as clarified in \cite{witten} gives rise to a very interesting possibility, namely to modify the nature of the factor, from being type $III$ into being type $II$ (or potentially type $I$ ) by properly regulating the UV infinite entanglement and by simultaneously adding {\it quantum gravity effects} \footnote{See the discussion on crossed products in the next chapter.}. We will come back to this point in a moment. Before doing that we will make some comments on the issue of causality or more generically how to fill the regions of the Penrose diagram not covered by the GNS representation.

\subsection{Emergent Causality}
The space-time interpretation of the GNS construction motivates the following natural question: 

{\it Can we define a new time, let us say $s$, and a new unitary generator $U(s)$ such that by flowing in this new time we can have:
\begin{equation}\label{causal}
\langle \Phi_0|O|\Phi_0\rangle - \langle b(s)|O|b(s)\rangle \neq 0
\end{equation}
for $s$ bigger than some critical value $s_0$ ?}

This is the question recently raised by Leutheusser and Liu \cite{LL1,LL2} in the context of the eternal AdS black hole. We will refer to the existence of such new time as {\it emergent causality}.

Physically the simplest naive answer to this question is easy to figure out. Namely the new time $s$ should be associated, in general, with a sort of generalized Bogolyubov transform mixing elements of ${\cal{A}}$ with elements in the commutant ${\cal{A}}'$. More precisely, let us imagine a localized perturbation in ${\cal{A}}$. We can think that this perturbation can be fully described using a sub algebra ${\cal{N}}$ of ${\cal{A}}$. In case we think in ${\cal{A}}$ as the algebra associated with a bounded region $E$ we will think the localized perturbation as defined by local operators with support in a subset of $E$. Now the emergent causality will be defined by means of a unitary transformation $U(s)$ and a new time $s$ such that by flowing in $s$ the sub algebra used to define the perturbation increases size i.e. $U(s){\cal{N}}U(s)^{-1} \supset {\cal{N}}$ until reaching a value $s_0$ at which $U(s_0){\cal{N}}U(s_0)^{-1} ={\cal{A}}$. For $s>s_0$ the desired emergent causality will appears if $U(s){\cal{N}}U(s)^{-1} \supset {\cal{A}}$ for $s>s_0$. It is in this regime $s>s_0$ when the operator $U(s)$ acts as a generalized Bogolyubov transform with elements $b(s)= U(s)bU(s)^{-1}$ for $b$ in ${\cal{N}}$ having non trivial projections in ${\cal{A}}'$. In other words if $U(s)$ satisfies the former conditions we will achieve (\ref{causal}) for $b$ a unitary element in ${\cal{N}}$.  

In principle we can make explicit this Bogolyubov transformation using a {\it purification} in the sense described in previous sections.

Recall that the purification associated with the GNS representation of the algebra ${\cal{A}}_R$ was defined using two commuting Heisenberg algebras
that we will denote now $a_k^{L,R},a_k^{\dagger, L,R}$. The creation operators were associated with a given energy $\omega(k)$ where this energy is conjugated to the global modular time $t$ \footnote{By defining $\omega(k)$ as the energy ( relative to modular time $t$ ) created by the operator $a_k^{\dagger}$ we are not assuming any form of well defined dispersion relation.}. Let us think geometrically the new time $s$ we are looking for as formally defining a Killing $\frac{\partial}{\partial s}$ not globally defined. In this case  we expect that flowing in $s$ will induce a non trivial Bolgolyubov transform of the type
\begin{equation}\label{Bog}
a^{L}_k(s) = A_{k,k'}(s) a_{k'}^{L} + B_{k,k'}(s) a_{k'}^{\dagger R}
\end{equation}
and similarly for $R$ generators with $a_k(s)= U(s)a_kU(s)^{-1}$. A priori we don't know if for the given GNS representation and the corresponding purification exits such unitary operator inducing a non trivial Bogolyubov transform. We will discuss in the next section under what conditions such unitary transformation exists. What we are presenting now are the conditions such operator needs to satisfy in order to define an emergent causal interplay between ${\cal{A}}$ and its commutant. In particular in the case ${\cal{A}}= {\cal{A}}_R$ the algebra of single trace operators associated with the $R$ exterior of the Penrose diagram of the AdS black hole, the emergent causal interplay between the $R$ and $L$ exterior regions will appear whenever the matrix $B$ in the Bogolyubov transform becomes non vanishing. 

Let us define the generator $G$ by $U(s)=e^{isG}$ and let us require $G$ to be bounded and selfadjoint operator i.e. the conditions needed to qualify as physical Hamiltonian. The former Bogolyubov transform is defining the operator $G$ at the level of the purification of the GNS construction. In this sense we can ask ourselves how this generator acts on the state $|\beta\rangle$. It is natural to impose that the action of $U(s)$ leaves the state $|\beta\rangle$ invariant. This means that $|\beta\rangle$ and $U(s)|\beta\rangle$ should be in the same GNS equivalence class i.e. $|\beta\rangle$ and $U(s)|\beta\rangle$ can only differ by the action of elements in the ideal $I_{|\beta\rangle}$. 

It is now obvious that for any mode $a_k^{\dagger R}$ representing a perturbation in the $R$ sector with modular energy $\omega(k)$ the evolution in $s$ time can be detected in the $L$ sector whenever $B(s)$ is non vanishing. Thus the existence of such Bogolyubov transform with $B$ non vanishing for some finite $s$ is the definition of an emergent causal interplay between the $R$ and $L$ sectors of the Penrose diagram. This Bogolyubov transform for the case of the eternal AdS black hole with the purification associated with the TFD state and the algebra of single trace operators has been nicely worked out in \cite{LL2}.

Specially interesting is when the Bogolyubov transform (\ref{Bog}) contains phase shift. By that we mean the case for which 
\begin{equation}
B_{k,k'}(s) = |B_{k,k'}|(s) e^{i\phi(k,k',s)}
\end{equation}
Physically these phase shifts describe the effective  scattering taking place when the mode flowing in $s$ time leaves the sector in which it was created. In the example of the eternal AdS black hole this phase shift contains information about the effect of the horizon when we cross it moving in $s$ time.

The Bogoliubov transform  (\ref{Bog}) defines generically the solution of the Heisenberg equation of motion
\begin{equation}
\frac{da_k^{L,R}}{ds} = i[G,a_k^{L,R}]
\end{equation}
Thus the existence of a phase shift implies that the generator $G$ can be decomposed as
\begin{equation}
G=G_0 +i G_1
\end{equation}

\subsection{Half sided modular inclusion}
The notion of half sided translation and half sided modular inclusion \footnote{For other uses of this notion see \cite{half1},\cite{half2},\cite{half3}.} naturally fits within the conditions on the emergent causality transformation $U(s)$ described above. First of all we formally defined a sub algebra ${\cal{N}}$ to host the perturbations. The representation of this sub algebra in the GNS Hilbert space will be done assuming the {\it same cyclic} state. Moreover we will assume that in this representation the sub algebra ${\cal{N}}$ defines a modular generator $K_{\cal{N}}$ as well as an anti linear map $J_{\cal{N}}$. We will say that ${\cal{N}}$ is half sided modular included in the original algebra ${\cal{A}}$ if
\begin{equation}\label{condition}
e^{itK_{\cal{A}}} {\cal{N}} e^{-itK_{\cal{A}}} \subset {\cal{N}}
\end{equation}
for $t \leq 0$. To get the physics meaning of this equation let us define a family of algebras ${\cal{N}}_t = e^{itK_{\cal{A}}} {\cal{N}} e^{-itK_{\cal{A}}}$. Then (\ref{condition}) implies that ${\cal{N}}_{t_1} \subset {\cal{N}}_{t_2}$ if $t_1< t_2$. Thus if we fix the algebra at {\it initial time} $t=0$ to be ${\cal{N}}$ what we get is that ${\cal{N}}_t \supset {\cal{N}}$ for $t>0$. This is the time evolution of the sub algebra generated by the modular transformations of the host algebra ${\cal{A}}$. Borchers theorem is telling us that in these conditions there exist a unitary transformation $U(s)$ with positive generator $G$ and a new time $s$ such that
\begin{equation}
e^{isG} {\cal{A}}e^{-isG} \subset {\cal{A}}
\end{equation}
for $s\leq 0$. The transformation $U(s)$ is a half sided translation of our original algebra ${\cal{A}}$. The key property of $U(s)$ is that for $s_0=-1$ we get
\begin{equation}
e^{is_0G} {\cal{A}}e^{-is_0G} = {\cal{N}}
\end{equation}
This means that in $s$ time the translated algebra ${\cal{A}}_s = e^{isG} {\cal{A}}e^{-isG}$ satisfies ${\cal{A}}_0 = {\cal{A}}$ and ${\cal{A}}_{-1} ={\cal{N}}$. The relation between modular time $t$ and the new time $s$ can be derived from
\begin{equation}\label{basic}
e^{iG s(t)} = \Delta_{{\cal{A}}}^{-it}\Delta_{{\cal{N}}}^{it}
\end{equation}
Using the standard BCH formula we get
\begin{equation}
e^{-it(K_{{\cal{A}}} - K_{{\cal{N}}}) - \frac{i^2t^2}{2} [K_{{\cal{A}}},K_{{\cal{N}}}] +..}
\end{equation}
Imposing $s(0)=0$ and defining $s(t)$ around $t=0$ by the Taylor series we easily get: $s(t) = e^{\alpha t}-1$ and 
\begin{equation}
G= \frac{K_{{\cal{N}}} - K_{{\cal{A}}}}{\alpha}
\end{equation}
and
\begin{equation}
G= \frac{-i}{\alpha^2} [K_{{\cal{A}}},K_{{\cal{N}}}]
\end{equation}
Moreover since  ${\cal{N}}$ is a sub algebra you can prove that
\begin{equation}
\Delta_{{\cal{N}}}\geq \Delta_{{\cal{A}}}
\end{equation}
which implies that $G$ defined as $K_{{\cal{N}}} - K_{{\cal{A}}}$ is positive.
Finally in order to fix the value of $\alpha$ above we use the basic relation \cite{Borchers}
\begin{equation}
e^{itK_{{\cal{A}}}} e^{isG}e^{-itK_{{\cal{A}}}} = e^{is(e^{-2\pi t})G}
\end{equation}
that sets $\alpha =2\pi$ and
\begin{equation}\label{time}
s(t) = e^{2\pi t}-1
\end{equation}

From the expression of the emergent time $s(t)$ we observe several interesting aspects. First of all the time $s_0=-1$ corresponds to $t=-\infty$. So evolving from time $t=-\infty$ to $t=0$ the sub algebra ${\cal{N}}$ becomes equal to the algebra ${\cal{A}}$. Moreover starting at $s=0$ and evolving using $U(s)$ for $s$ negative we reduce the algebra ${\cal{A}}$ until reaching ${\cal{N}}$ while evolving with $U(s)$ for $s$ positive and starting at $s_0=-1$ we first increase ${\cal{N}}$ into ${\cal{A}}$ for a positive time $s$ equal one  and we further increase ${\cal{N}}$ beyond ${\cal{A}}$ for positive $s$ bigger than one. In summary from $t=-\infty$ to $t=0$ we increase the region where the initial perturbation is localized while remaining inside the region $E$ used to define the algebra ${\cal{A}}$. When we continue from time $t=0$ to $t=\infty$ the initial perturbation effectively leaves the region $E$ and causal interchange starts. We need to be careful with the heuristic picture since what we are really showing is that the evolution in $s$ of the initial perturbation after some time $s=1$ starts to contribute to (\ref{causal}). It is also worth to stress the formal similarity between the emergent time defined by (\ref{time}) and a typical expanding time \footnote{The relation between the emergent time $s$ and $t$ in (\ref{time}) is reminiscent of the basic GR relation in the black hole geometry between the coordinate $u$ vanishing at the horizon and the time $t$, namely $u= e^{-\frac{t}{4GM}}$.}.

\subsubsection{Small digression}
The crucial aspect of the former construction is the non commutativity of the modular generators $K_{{\cal{N}}}$ and  $K_{{\cal{A}}}$. Heuristically this lack of commutativity gives us the clue on how the emergence of causality takes place. Indeed if we think in a perturbation created by some element in the sub algebra ${\cal{N}}$ with a well defined value of the modular energy $
K_{{\cal{N}}}$ the lack of commutativity and the definition of $G$ implies the existence, for this perturbation, of some non vanishing variance $\Delta(G^2)$ i.e. a non vanishing fluctuation for the "$G$-energy". It is this fluctuating energy the one you can use to inspect the region, in the case of the eternal AdS black hole, in the black hole interior. After coupling to gravity i.e. when you include $frac{1}{N}$ corrections for $G_N= \frac{1}{N^2}$ with $N$ defined by the dual CFT, these fluctuations will induce a non vanishing gravitational potential on the horizon. In the last chapter we will extend this line of reasoning to the context of Cosmology. 

An important preliminary question is how to measure the physical consequences of the non commutativity between the modular generator $K$ and the emergent time generator $G$. The most natural way to unveil the relations between the generator $G$ and the modular Hamiltonian $K_{{\cal{A}}}$ is to consider the family of operators
\begin{equation}\label{one}
\rho_{f}(s) = U(s) \rho_{f} U(s)^{-1}
\end{equation}
where $\rho_f = e^{K_{{\cal{A}}}}$ and where $f$ indicates the state used in the GNS representation of ${\cal{A}}$. A quantum estimation theory question is how much information we have, in order to estimate the value of $s$, if we use as data the expectation values, defined by $f$, of any set of observables. The invariance of the cyclic state defined by $f$ under the action of $U(s)$ implies that
\begin{equation}
f(a(s))= f(a)
\end{equation}
so on the basis of the collection of {\it data} defined by the expectation values $f(a)$ for $a$ any element in ${\cal{A}}$ we cannot extract any information to estimate the value of $s$. This contrast with the familiar situation in quantum estimation theory where, if we have a density matrix depending on a parameter in the form given by (\ref{one}), we can define the information about $s$ using the quantum Fisher information \cite{Paris}. In particular for a density matrix satisfying 
$\frac{\partial \rho}{\partial s} = \frac{1}{2}(L\rho + \rho L)$ the quantum Fisher $I_F(\rho)$ is given by
 \begin{equation}
 I_F(\rho) = \frac{1}{4}Tr(L^2\rho)
 \end{equation}
 An equivalent quantity is defined by the Wigner-Yanase information \cite{WY}. For a density matrix $\rho$ and a self adjoint operator $G$ characterizing the dependence of $\rho$ on the parameter to be estimated the WY information  is defined by:
 \begin{equation}
 I_W(\rho) = - Tr ([\rho^{1/2},G]^2)
 \end{equation}
 Obviously these definitions require to have a state i.e. a linear form $f$ satisfying the trace property. This is not the case for type III factors. In summary we find an interesting situation where $\rho_f(s)$ defined by (\ref{one}) changes with $s$ if $[K_{\cal{A}},G]$ is non vanishing but where the absence of trace prevents us to use the standard quantum Fisher information to estimate the value of the parameter $s$. How to generalize to type III factors the notion of quantum Fisher information for the family of states defined by (\ref{one}) ? This can be answered using the measure introduced by Connes and Strormer \cite{cS}. 
In our case since we are interested in estimating $s$ we will define
 \begin{equation}
 I(f,x,s) = ||( Jx^{+}(s)J - x(s))||_f
 \end{equation}
 where we denote $||...||_f$ the norm defined by the state $f$ used to define the cyclic and separating state in the GNS construction and where $x(s)$ is defined as $x(s) =U(s)xU(s)^{-1}$ for any $x$ in ${\cal{A}}$. The reason this quantity is interesting is because, thanks to Tomita relations (\ref{tomi1}) and (\ref{tomi2}), and the assumed invariance of the cyclic state, it captures the non commutativity of $K$, defining the modular generator, and $G$. 

Using a purification defined by the state $|\beta\rangle$ we can define the GNS state $|[a_k^{R,\dagger}(s)]\rangle$ where the equivalence class is defined relative to the ideal $I_{|\beta\rangle}$. We are now interested in estimating how this path in ${\cal{H}}^{GNS}$ depends on $s$. The quantum variance $\Delta(G^2)$ for this state will be formally defined by:
 \begin{equation}
 I^{GNS}(a_k^{R,\dagger}, \beta,s) = \langle[a_k^{R,\dagger}(s)] |G^2 |[a_k^{R,\dagger}(s)]\rangle - (\langle[a_k^{R,\dagger}(s)] |G |[a_k^{R,\dagger}(s)]\rangle)^2
 \end{equation}

\section{Time and the Crossed Product}
Let us start introducing the notion of crossed product of a von Neumann algebra ${\cal{A}}$ by a group action \cite{Phillips}. We will consider the group as representing time translations and the action of the group on the algebra as defined by a map $\alpha:t \rightarrow Aut ({\cal{A}})$ from the group, that we parametrize by $t$, into automorphisms of the algebra. Now we define square integrable functions from the group into the algebra i.e.  functions $a: t \rightarrow {\cal{A}}$. The definition of the crossed product will consist in promoting this space of functions $a(t)$, valued in the algebra, into a well defined von Neumann algebra. 

Generically for an algebra ${\cal{A}}$ a group $G$ and an action of $G$ on ${\cal{A}}$ defined by the automorphisms $\alpha(g)$, the crossed product is represented as
\begin{equation}
{\cal{A}}\rtimes_{\alpha} G
\end{equation}
A special case is when $G$ represents the group of time translations, ${\cal{A}}$ is a type $III$ factor and we use as $\alpha$ the modular automorphism i.e. $\alpha_t(a) = \Delta ^{it} a \Delta^{-it}$ for $a$ any element in ${\cal{A}}$. In this case the crossed product will depend on the particular cyclic and separating state used to define the modular automorphism $\Delta$ \footnote{For a discussion of the dependence on the cyclic state, based on Connes results \cite{Connes}, see \cite{witten}.}. 

For $G$ the group of time translations the crossed product algebra ${\cal{A}}\rtimes_{\alpha} G$ is the $C^*$ algebra of continuous functions with compact support $a:t \rightarrow {\cal{A}}$ with multiplication defined by the following convolution
\begin{equation}
ab(t) = \int dt' a(t') \alpha_{t'}(b(t-t'))
\end{equation}
In the case of the group of time translations the adjoint $a^*(t)$ is defined by 
\begin{equation}
a^*(t) = \alpha_t(a(-t)^*)
\end{equation}
For generic group $G$ equipped with a Haar measure the adjoint will be defined using the corresponding modular function \cite{Phillips}. In the case of interest of time translations the modular function reduces to one.

Intuitively if ${\cal{A}}$ is the algebra of local observables associated with a bounded region of space-time and we can define a geometric notion of time translation on this region, that acts on the algebra as an automorphism, the elements in the crossed product are just continuous {\it paths} in ${\cal{A}}$ with compact support. 

A representation of this crossed product algebra on a Hilbert space ${\cal{H}}$ is defined using two ingredients. On one side the representation $\pi$ of ${\cal{A}}$ 
in the space of bounded operators ${\cal{B}}({\cal{H}})$ of ${\cal{H}}$ and on the other side the representation $v$ ( in our case) of the time translation group in terms of a {\it unitary} operator acting on ${\cal{H}}$.
We can define this representation as
\begin{equation}
v(t) = e^{it\hat X}
\end{equation}
with $\hat X$ a unitary operator. Generically we should require the {\it covariance condition} namely
\begin{equation}\label{covariance}
e^{it\hat X} \pi(a) e^{-it \hat X} = \pi(\alpha_t(a))
\end{equation}
Note that this covariance condition puts in contact the particular $\alpha_t$ used to act on the {\it algebra} and the operator $\hat X$ used to represent the group of time translations on the Hilbert space. 

Next we need to define the {\it representation} of the crossed product algebra. This means we must associate with a given path $a(t)$ in the crossed product a bounded operator in ${\cal{B}}({\cal{H}})$. The action of the bounded operator associated to the path $a(t)$ on a generic state $|\psi\rangle$ in the Hilbert space is defined by

\begin{equation}
\int dt \hat a(t) e^{it\hat X}|\psi\rangle
\end{equation}
where by $\hat a(t)$ we mean the representation $\pi$ of $a(t)$ in terms of bounded operators of the Hilbert space and where $\hat X$ is the unitary operator defined above.

The expectation values of the elements in the crossed product algebra ( that for time translations we will simply denote by ${\cal{A}}\rtimes R$ ) are defined by
\begin{equation}\label{key}
\langle a(t) \rangle = \langle \psi|\int dt \hat a(t) e^{it\hat X}|\psi\rangle
\end{equation}

The final step in the characterization of the crossed product is to identify the Hilbert space where we are defining its representation. Imagine that you have defined a GNS representation of the algebra ${\cal{A}}$ and denote as usual this Hilbert space as ${\cal{H}}_{GNS}$. Now you are interested in defining a new Hilbert space where to represent the crossed product of ${\cal{A}}$ by the group of time translations. This Hilbert space can be identified as {\it the space of square integrable functions $L^2(R,{\cal{H}}_{GNS})$}. In other words, the states representing the crossed product algebra are "paths" in the Hilbert space defining the GNS representation of ${\cal{A}}$. We can formally represent these paths as $|\psi \rangle(t)$ and to define the scalar product in 
$L^2(R,{\cal{H}}_{GNS})$ as
\begin{equation}
 \int dt \langle \psi (t)|\phi(t)\rangle
\end{equation}
Now for a path $|\psi\rangle (t)$ in $L^2(R,{\cal{H}}_{GNS})$ covariance implies
\begin{equation}
e^{it_1\hat X} |\psi\rangle (t) = |\psi\rangle (t-t_1)
\end{equation}

\subsection{Heuristic characterization of crossed products}
The reason the crossed products of a type $III$ factor by the group of time translations can be crucial to deal with the UV divergences of entanglement can be understood as follows. One of the consequences of the divergent entanglement for type $III$ factors is that the modular automorphism defining time translations is an {\it outer} automorphism. In that sense you can associate infinite entanglement to what we can poetically call {\it outer time}. The crossed product is a mathematical construction that has as main goal to define an extended algebra where the time translation generator is included. In other words the extended algebra transforms the original {\it outer time} into what we can denote an {\it inner time}. This is done going from the original algebra to the algebra of paths and changing the Hilbert space representation from the original GNS Hilbert space into the space of paths $L^2(R,{\cal{H}}_{GNS})$. The reason this is a crucial step to understand the way to give meaning to finite entanglement or generically to finite entropy is because since the new extended algebra contains {\it time translations} as inner automorphism must necessarily be either type $I$ or type $II$ having a well defined notion of quantum micro states or at least of a density matrix. 

Of course if you are working with an {\it inner time} from the beginning to use the space of paths $a(t)$ in the algebra or equivalently the space of paths to define the Hilbert space is not producing anything new.

\subsection{Quantum Variance}
After we count with the former heuristic interpretation of the crossed product we can gain some physics intuition visualizing the elements in $L^2(R,{\cal{H}}_{GNS})$ where we are representing our crossed products as paths in the GNS Hilbert space representation $|\psi\rangle(t)$. A special sort of paths are those where {\it all states in ${\cal{H}}_{GNS}$ defining the path are indistinguishable by performing measurements associated with observables in ${\cal{A}}$ }. Physically we can think of these trivial paths as associated with a conserved value of {\it energy}, or more precisely as associated to eigenstates of $\hat X$. On those paths the crossed product trivializes and $\hat X$ is acting as a central element. 

Based on this intuition we can define a {\it quantum variance} for $\hat X$ as follows. Take an element $\psi$ in $L^2(R,{\cal{H}}_{GNS})$ defined by a path $|\psi\rangle (t)$. Now for arbitrary value $t_0$ we can define
\begin{equation}
I_F(t_0; |\psi) = \Delta (\hat X^2; \psi,t_0)
\end{equation}
where $\Delta (\hat X^2; \psi,t_0)$ is formally defined as $\langle \psi(t_0)|\hat X^2 |\psi(t_0)\rangle - (\langle \psi(t_0)|\hat X |\psi(t_0)\rangle)^2$ where $\hat X$ is an {\it outer} automorphism of the original algebra ${\cal{A}}$.

We can formally think $I_F$ as defining the quantum Fisher metric on the GNS Hilbert space. The physical meaning of this quantity is that if {\it non vanishing} is telling us that the elements in the path defining the state, in the extended Hilbert space $L^2(R,{\cal{H}}_{GNS})$, are quantum mechanically distinguishable.

Note that if the operator $\hat X$ commutes with all the elements in ${\cal{A}}$ it effectively means, due to the covariance condition (\ref{covariance}) that $\alpha_t$ defining the crossed product is trivial. This already implies that we cannot define this $\alpha_t$ using the modular automorphisms $\Delta^{it}$ of the algebra ${\cal{A}}$. This trivial case simply corresponds to identify some central conserved charge. Obviously in this case the defined quantum variance is zero and the Hilbert space of paths is simply the Hilbert space where we have defined the GNS representation. In this trivial case the formal crossed product is not a factor since, by construction, it contains a non trivial center. 
\subsubsection{Quantum gravity corrections}
An extremely interesting question raised in \cite{witten} is if {\it quantum gravity corrections} can be organized in the form of a non trivial crossed product defining a good factor. In order to understand the underlying logic let us think in the case in which we can identify geometrically a well defined time translation transformation but that {\it due to UV divergences} you cannot represent this transformation in $Aut{\cal{A}}$. Intuitively you could hope that by coupling to gravity, more precisely by including quantum gravity corrections, you can regulate these UV divergences, and give a precise meaning to the corresponding group of time translations using well defined elements in $Aut{\cal{A}}$. 

To fix ideas let us consider as a concrete example the case of the eternal AdS black hole worked out in \cite{witten}. The QFT type $III$ algebra ${\cal{A}}$ is associated with the left and right external pieces of the Penrose diagram. More precisely, as already discussed, we associate ${\cal{A}}$ to the right side and the commutant ${\cal{A}}'$ to the left side. Imagine you are the right observer living in the right external side of the black hole. A time translation group is defined by the global time translations acting {\it only in the right side}. The generator of this time translations is what we have denoted normally $H_R$. As stressed due to the infinite entanglement $H_R$ is not acting as a well defined operator on states in the GNS representation Hilbert space associated with ${\cal{A}}$. This is a case where you identify a classical group of time translations {\it but} you cannot define the corresponding representation $\alpha_t$ in $Aut{\cal{A}}$. However you can {\it regulate} gravitationally this quantity. In the case the holographic representation is available you can codify the divergences of $H_R$ in the large $N$ dependence with $N$ defined by the holographic representation i.e. $G_N \sim \frac{1}{N^2}$. Thus you can define the regulated $H_R$ as $\frac{H_R}{N}$. This regulated generator will be order $N^0$ in powers of $N$ or in more physical terms will be order $M_P^0$. Now you can build the crossed product using as the group of transformations the time translations on the right side generated by {\it the regulated} $H_R$. This regulated $H_R$ is however {\it central} and its value can be interpreted as the classical black hole mass, let us say $M_{BH}$. We can, using the former notation, to identify the regulated $H_R$ as a central $\hat X_0$ operator. 

Thus the regulated $H_R$ leads to a trivial crossed product. Can we promote this trivial crossed product into something non trivial including the quantum gravity corrections ?

In essence what we are looking for is a quantum gravity perturbative series for the operator $\hat X$ of the type
\begin{equation}
\hat X = \hat X^{0} M_P^{0} + \hat X^{1} (M_P)^{-1} +...
\end{equation}
with $\hat X_0$ central. Using now the covariance condition (\ref{covariance}) we can easily imagine the simplest form of $\hat X^{1}$ namely

\begin{equation}\label{correction}
\hat X^1 = K \hat X_0^2
\end{equation}
where we keep the generator $\hat X$ as having energy dimension and where $K$ is defined as usual as the generator of the modular automorphism $\Delta^{it}$. 

Now we can define a crossed product with this $\hat X$ and with $\alpha_t$ the one naturally defined by the modular automorphism of the algebra ${\cal{A}}$. Let us denote this crossed product ${\cal{A}}\rtimes_{\alpha} R$. In other words we can use the covariance condition to define $\alpha_t$ once we have defined a regulated and quantum corrected $\hat X$ operator. 


\subsection{Back to variance}
Let us now focus on the Hilbert space $L^2(R,{\cal{H}}_{GNS})$ on which we are representing the crossed product ${\cal{A}}\rtimes_{\alpha} R$. As discussed elements in this Hilbert space are "paths" parametrized by time $t$ in the GNS Hilbert space. Depending on the nature of ${\cal{A}}\rtimes_{\alpha} R$ as a factor we can find in this Hilbert space "paths" representing a well defined density matrix in case the crossed product is a factor of type $II$. 

What kind of factor is the so defined crossed product is in general not an easy question. The approach used in \cite{Witten1} and \cite{Witten2} is roughly the following. Take special states in $L^2(R,{\cal{H}}_{GNS})$ of the type $|\Psi\rangle f(t)$ with $f(t)$ in $L^2(R)$ and where the state $|\Psi\rangle$ is {\it independent} of $t$. These states are denoted {\it classical-quantum states} in \cite{witten}. Assume that for some $f(t)$ and for $|\Psi\rangle$ the cyclic state in ${\cal{H}}_{GNS}$, the so defined state is cyclic and separating in $L^2(R,{\cal{H}}_{GNS})$. Now using the standard approach define the modular automorphism associated to this state. If the corresponding modular generator is in the crossed product algebra then the crossed product is necessarily of type $I$ or type $II$. If it is of type $II$ we can have two options, namely type $II_1$ or type $II_{\infty}$. In both cases we can define a trace and in principle to associate with a given state in $L^2(R,{\cal{H}}_{GNS})$ a density matrix. 

Let us formally denote for a given state $|\psi\rangle$ in $L^2(R,{\cal{H}}_{GNS})$, that recall is defined by a path of states in the GNS Hilbert space, a density matrix $\rho_{\psi}$. This will allows us to associate with this state the corresponding von Neumann entropy $S(\rho_{\psi})$. In order to do that we need to define a trace such that $tr(\rho_{\psi} a) = \langle \psi|a|\psi\rangle$ for  $|\psi\rangle$ in $L^2(R,{\cal{H}}_{GNS})$ and $a$ a generic element in the crossed product. However  if we consider generic states $|\psi\rangle(t)$ i.e. states of , for instance, the type $|\Psi\rangle(t). f(t)$ with $|\Psi\rangle$ {\it depending} on $t$ we can equally associate with the path in $L^2(R,{\cal{H}}_{GNS})$ a pure {\it quantum information} besides the von Neumann entropy, namely the quantum variance $I_F(|\Psi\rangle)$ defined above. This quantum variance can be defined in the standard way as quantum Fisher \cite{Paris} or quantum Wigner-Yanase information \cite{WY}. Indeed if $|\psi\rangle(t)$ represents a density matrix $\rho$ this information is given, provided we have a crossed product of type $II$ with a trace, by
\begin{equation}
I_F\sim  tr(\rho \hat X^2)
\end{equation}
that measures the lack of commutativity between $\rho$ and $\hat X$. 

Before going on it could be worth to make some comments. First of all the natural candidate to define a density matrix associated with a cyclic and separating state {\it in $L^2(R,{\cal{H}}_{GNS})$ } will be the corresponding Tomita operator for this cyclic state relative to the crossed product algebra. This is indeed the candidate used in \cite{witten2}. Moreover in order to define the trace we should use the prescription defined in (\ref{key}) for the crossed product and the KMS property needed to achieve the trace property $tr(ab)=tr(ba)$. 

Finally note that in our definition of crossed product the integral in (\ref{key}) is defined using a Haar measure on the group $G$ we are using to define the crossed product. In case we identify this group with time translations we represent the elements in $G$ by the time $t$ of the time translation. Thus in principle the value of $t$ will go from $-\infty$ to $+\infty$. 

In $L^2(R,{\cal{H}}_{GNS})$ we can try to distinguish different states and to define a distinguishability distance between them. That can be done using Araki's operators associated with the states in $L^2(R,{\cal{H}}_{GNS})$. The quantum information we are defining is {\it not} this "metric" in $L^2(R,{\cal{H}}_{GNS})$ but the {\it microscopic} distinguishability between the elements composing the path $|\psi\rangle(t)$ in $L^2(R,{\cal{H}}_{GNS})$. 

Concerning the question on entropies let us recall that one of the main targets of \cite{witten} ( in the opinion of the author ) is to use crossed products to frame in rigorous terms the approach of  Susskind and Uglum \cite{Susskind} to Bekenstein generalized entropy \cite{Bekenstein}. This is done using the two well defined operators for the crossed product, namely the modular operator $K$ and the operator $\hat X$, to identify the cancellation between the divergences of the radiation entropy and the ones of $\frac{A}{4G_N}$.

\subsection{Some comments on the case of de Sitter }

A similar construction to the one used for the eternal black hole can be developed for de Sitter \cite{witten2}. In this case the analog of the right and left external sides of the eternal black hole are the two {\it static patches}. Thus we can associate a type $III$ factor to one static patch and the commutant to the other. If we denote ${\cal{A}}_{dS}$ the type $III$ algebra associated with the "right" static patch i.e. the one corresponding to the right observer, the commutant ${\cal{A}}'_{dS}$ can be identified with the algebra of local observables on the left static patch. The GNS representation of ${\cal{A}}_{dS}$ defines the Hilbert space ${\cal{H}}^{dS}_{GNS}$ provided we choose a cyclic and separating state. Let us assume that the well known Bunch Davis (BD) vacuum \cite{BD}, to be denoted as $|BD\rangle$, defines such cyclic and separating state in ${\cal{H}}^{dS}_{GNS}$ \footnote{This state is characterized by the euclidean continuation. However there are other de Sitter invariant states that can be formally defined using the Mottola-Allen transformation \cite{Motola}}. We can think of the standard BD vacuum as a {\it purification} of the cyclic and separating state defining the GNS representation in the same way we introduced the TFD state. Once the GNS representation is given we have defined the modular automorphisms $\Delta_{dS}^{it}$ with generator the modular Hamiltonian $K_{dS}$. Geometrically $K_{dS}$ is associated with a Killing vector that defines time translations on one static patch and the corresponding {\it time reversal} on the other static patch. The KMS property of the  $|BD\rangle$ allows us to define the de Sitter temperature $\beta_{dS}$ and a Hamiltonian associated to $K_{dS}$ in the standard way as $\beta_{dS}H_{dS}= K_{dS}$. 

Now, as we did in the case of the black hole, we can consider time translations acting {\it only on one static patch} let us say the right patch. These are the time translations for the right observer and these are the ones that have the physical meaning for her of positive energy. Let us denote this generator $H_R$. As already discussed the UV divergent entanglement implicit in the type $III$ factor we use to describe the observables on the static patch has as a consequence that $H_R$ is not well defined as operator acting on the Hilbert space ${\cal{H}}^{dS}_{GNS}$. 

As pointed out in \cite{witten2} in the moment we couple to gravity and due to the fact that the static patch goes from $r=0$ to the cosmological horizon we need to implement the time translations as a {\it gauge symmetry}. At the level of the full GNS Hilbert space we could try to implement this constraint using the modular Hamiltonian as the generator of the gauge symmetry. If we do that we will be in trouble since the modular action is {\it ergodic} and essentially the so defined invariant part of ${\cal{A}}_{dS}$ will be trivial. Intuitively this means that the orbit in ${\cal{A}}_{dS}$ generated by the modular automorphisms is, due to ergodicity, dense and therefore the space of orbits is trivial.

However we can follow the path designed in the definition of the crossed product where now we are interested in the action of time translations acting {\it only} on the static patch of one observer. 

Following the former discussion this crossed product will be defined in terms of $K_{dS}$ and $\hat X_{dS}$ with $K_{dS}$ the modular generator and $\hat X_{dS}$ the unitary operator implementing unitarily time translations in the Hilbert space representation of the crossed product. Doing that we will reproduce the former discussion. 

However we can slightly change our approach to crossed products including not only the operator $\hat X$ introduced above {\it but also} a canonical conjugated operator that we will denote $\hat T$ and that we can think somehow as a {\it clock operator}. In essence what we are adding is an operator $\hat T$ such that $[\hat X,\hat T]= i\hbar$. Once we have introduced the clock operator $\hat T$ the operator $\hat X$ can be interpreted as representing the Hamiltonian of the observer that is measuring with this clock. Now we can use the {\it dual energy version} of the crossed product where instead of using functions $t: \rightarrow {\cal{A}}$ we use functions $u:\rightarrow {\cal{A}}$ with $u$ in the spectrum of $\hat X$. In more simple terms now you crossed by the action of changes in $u$ with unitary operator the clock operator $\hat T$.

Now on the basis of requiring {\it positivity} of $\hat X$, the natural Hilbert space representing the crossed product will be
\begin{equation}\label{Hilbert}
L^2( R^+, {\cal{H}}^{dS}_{GNS})
\end{equation}
An intriguing possibility that we will briefly discuss in the next chapter is to identify $R^+$ in (\ref{Hilbert}) as the half sided range of an {\it emergent time}.

In connection with this last comment recall that in the previous chapter we have discussed, following \cite{LL1},\cite{LL2} the introduction of new emergent times using Borchers half sided modular inclusion. After our qualitative discussion on the relation between emergent times and Bogolyubov transformations ( see section II (F and G)) we could suggest as the natural setup for defining emergent times the crossed product

\begin{equation}
{\cal{A}}\rtimes_{\alpha(G)} R^{+}
\end{equation}
where we reduce the emergent time to be positive on the basis of the half sided inclusion. The potential role of this form of crossed product with respect to an emergent time will be considered in the next section where we will try to frame inflationary cosmology as an example of crossed product.

\subsubsection{Clocks and Observers}
In the previous section we have defined a {\it clock algebra} considering in addition to $\hat X$ the conjugated variable $\hat T$. From the point of view of the crossed product algebra we can think of $\hat T$ as an outer automorphism and to define a so called double crossed product to add the generator $\hat T$. This indeed can be done if we don't insist in interpreting $\hat X$ as associated with a physical positive energy. If we do that, as it is well known in quantum mechanics, the time or clock operator $\hat T$ is not defined. Let us ignore, for a moment, this issue and think of a clock Hilbert space as a representation of the clock algebra generated by $\hat X$ and $\hat T$. We could formally think in wave functions $f(u)$ describing a state in this Hilbert space with $f(u)$ the probability amplitude to have energy $u$. These should be real functions square integrable and normalized to one
\begin{equation}
\int_0^{\infty} |f(u)|^2 =1
\end{equation}
We can think of $\hat X - K$, with $K$ the modular Hamiltonian and $\hat T$ as the algebra describing the observer equipped with a clock $\hat T$ with Hamiltonian $\hat X - K$. 

In reality once we take into account that $\hat T$ is not defined for $\hat X$ representing a physical and positive energy we should think of $\hat T$ in terms of a quantum time estimator.

Now the so called {\it classical-quantum states} in the crossed product will be of the type 
\begin{equation}
|\Phi\rangle. f(u)
\end{equation}
with $|\Phi\rangle$ in ${\cal{H}}_{GNS}^{dS}$. In these states the state $|\Phi\rangle$ describing dS is assumed to be independent of $u$. Intuitively we can think of $f(u)$ as probability distributions.

The importance of these states is that for them the uncertainties implicit in the form of $f(u)$ are not creating any gravitational back reaction on the state $|\Phi\rangle$ we properly associate with the dS static patch, recall that in these states $|\Phi\rangle$ is independent of $u$.

In \cite{witten2} some concrete examples of classical quantum states are described. In those states, that we will discuss in a moment, we find the natural uncertainties in the values registered by the clock. However these uncertainties are pure clock properties and fully independent of the time dependence of the underlying de Sitter metric. Certainly we should, in order to define a reliable clock, to assume that these intrinsic clock uncertainties are much smaller that the typical change of the expanding metric. 

Let us now pause to discuss this issue with a bit more detail. As has been pointed out for several authors \cite{Banks1},\cite{Banks2} once we include the clock in the static patch we reduce the entropy in an amount determined by the clock energy. Moreover any physical fluctuation created in the static patch ends, after some time, leaving the static patch that at the end becomes associated with the maximal entropy state. However to assign entropy we need to move out from the type $III$ factor and to describe this situation in the corresponding type $II$ crossed product. The maximal entropy state in this crossed product is associated with a state of type $|\Phi_0\rangle f(u)$ with $|\Phi_0\rangle$ a cyclic state with $K_{dS}|\Phi_0\rangle =0$ and with $f(u) = e^{\beta_{dS}u}$ for $\beta_{dS}$ ( the Hawking Gibbons temperature) the KMS temperature associated with $|\Phi_0\rangle$.

\section{Cosmology as a crossed product}
Inflationary Cosmology is the most accepted paradigm to solve the classical problems raised by Big Bang Cosmology as it is the horizon or flatness problem \footnote{For a good general review see \cite{Baumann}.}. Cosmology is generically described in terms of a  FRW metric accounting for the history of the Universe. In conformal time coordinates, this metric is given by
\begin{equation}
ds^2 = a^2(\eta)(-d\eta^2 + dx^2)
\end{equation}
for $\eta$ the conformal time defined by $d\eta=\frac{dt}{a(t)}$ for $t$ the physical time.  The key idea of inflationary Cosmology is to extend the conformal time axis from $-\infty$ to $+\infty$ with the inflationary period going from some initial time $\eta_{in}$ ( potentially equal to $-\infty$ ) until $\eta=0$ where $a(\eta)$ is of dS type i.e. exponentially expanding and with a change of regime at the end of inflation $\eta\sim0$ to a matter dominated or radiation dominated Universe. For our present discussion it will not be important to consider the details of recombination or the precise time defining the last scattering surface. In most models the inflationary period is described in terms of a quasi de Sitter phase characterized by the slow roll parameters defining the underlying inflaton potential. 

In this scenery the primordial quantum fluctuations that leave the cosmological horizon during the inflationary period reenter in the post inflationary epoch leading to the observable features of the CMB spectrum. A simplified working assumption is that the features of the power spectrum at reentering time are the same as the primordial features defining the scalar curvature power spectrum at horizon exit in the primordial inflationary period. In general is assumed that this primordial power spectrum encodes the information about the slow roll parameters defining the primordial quasi de Sitter phase.

In what follows we will start to rethink the former picture from a purely {\it quantum mechanical } point of view based on the discussion of the crossed product description of the Hilbert space we associate to quantum de Sitter, namely $L^2(R^+,{\cal{H}}_{dS})$.

The simplest possibility will be to associate the primordial inflationary period with a quantum state in $L^2(R^+,{\cal{H}}_{dS})$ with $R^+$ representing the conformal time in the inflationary period. Thus our basic definition for a quantum description of inflationary cosmology will be:

{\it To describe this period by a quantum state in $L^2(R^+,{\cal{H}}_{dS})$ with $R^+$ representing the conformal time during the inflationary period.}

The special characteristic of this state will be to be a purely quantum state and not what we have described in the former section as a classical-quantum state. More precisely the inflationary cosmology should be described by some state $|\Phi_{cos}\rangle (\eta)$ that cannot be factorized in the form $|\Phi\rangle.f(\eta)$ for some $\eta$ independent $|\Phi\rangle$.

The next ingredient defining this quantum mechanical crossed product version of Cosmology will be to focus not in the von Neumann entropy that formally we can associate with $|\Phi_{cos}\rangle(\eta)$ by defining the corresponding density matrix, using  the type $II$ trace, but, instead, on the quantum Fisher information associated with the variance of $\hat X$. As discussed in the previous section this quantum Fisher information is telling us about the quantum distinguishability of the different states in ${\cal{H}}_{dS}$ contributing to the "path" $|\Phi_{cos}\rangle(\eta)$.

It is the quantum variance of $\hat X$ for the state $|\Phi_{cos}\rangle(\eta)$ the one we could try to use to make contact with the experimental features of inflation. In very formal terms the relation we will try to develop can be summarized as:

\begin{equation}\label{keys}
\Delta (\hat X^2; |\Phi_{cos}\rangle(\eta)) \Longleftrightarrow {\cal{P}}
\end{equation}
for ${\cal{P}}$ representing the primordial power spectrum of scalar curvature fluctuations.

Note the drastic change of point of view we are suggesting. Instead of thinking in different, and potentially infinite distinct types of inflaton potential, we want to use quantum states in the crossed product Hilbert space to describe the inflationary epoch. The {\it classical} quasi de Sitter "history" is now converted into a particular quantum state $|\Phi_{cos}\rangle(\eta)$ in the de Sitter crossed product. The non trivial features of this path or equivalently of the primordial "history" are now translated into the quantum distinguishability of the different states contributing to the "path" $|\Phi_{cos}\rangle(\eta)$. That is what we try to catch quantitatively in terms of the quantum variance of $\hat X$ on this state and hopefully to derive from that some precise and model independent information about the observable curvature power spectrum.

Probably the reader will be asking herself what happens in this algebraic scheme when inflation ends. The simplest view is to describe the post inflationary epoch, at least in first approximation and assuming we live at present in a state of exponential expansion, by another crossed product Hilbert space where now $R^+$ represents the conformal time in the post inflationary epoch. This will leave us to think in some different state $|\Phi_{post}\rangle(eta)$ living in the post-inflationary crossed product Hilbert space. Most likely the key difference between $|\Phi_{cos}\rangle(\eta)$ and 
$|\Phi_{post}\rangle(eta)$ could be that $|\Phi_{post}\rangle(eta)$ is expected to be a classial-quantum state of the type discussed in the previous section. The quantum description of the transition, the inflationary Big Bang, should be described in a bigger Hilbert space about which we have, at this moment, nothing intelligent to say.

\subsection{Brief recap of the theory of primordial quantum fluctuations}
To start with let us recall that geometrically the inflationary period can be described using the {\it planar patch} of de Sitter space-time. In the Penrose diagram this planar patch contains two pieces. One piece is the static patch $E$ of either the north (or south) observer with vertical boundary at $r=0$. The other piece is the upper (lower) triangular region with horizontal boundary at $r=\infty$. The boundary of the static patch corresponds to the cosmological horizon and the upper triangular side to the exterior. The full planar patch can be foliated using the conformal time coordinate. In the case of a pure de Sitter period this conformal time is defined by $e^{H_bt} = -\frac{1}{\eta H_b}$ for $H_b$ the Hubble constant.

Probably the most important prediction of inflationary Cosmology is the power spectrum of primordial fluctuations. The derivation of this primordial spectrum is based on the equation of small {\it quantum fluctuations} in a space-time background defined by a matter content with equation of state characterized by
\begin{equation}\label{state}
\epsilon_{cl} = \frac{3}{2} (\frac{\rho+p}{\rho})
\end{equation}
with $\rho$ and $p$ fully determined by the classical inflaton dynamics. More precisely you define the most general metric with linear scalar perturbations and identify the gauge invariant gravitational potential $\Phi$ associated to the scalar fluctuation. The case $\epsilon_{cl}=0$ corresponding to a pure cosmological constant leads to de Sitter space-time. For $\epsilon_{cl}$ small we get quasi de Sitter space time.  In local planar coordinates you can define the different Fourier modes $\Phi_k(\eta)$ and to identify $k$ with a comoving momentum. 

In understanding primordial quantum fluctuations it is convenient to distinguish pure QFT defined on the planar patch in the limit where {\it quantum gravity} effects are decoupled from the first quantum gravity correction. As already discussed when a holographic dual representation is available these quantum gravity effects are order $1/N$ for $N$ characterizing the CFT dual with $G_N \sim \frac{1}{N^2}$. In the QFT regime we have effectively $M_P=\infty$ ( $N=\infty$ ) and we can parametrize the small classical departure from pure de Sitter in terms of $\epsilon_{cl}$. Quantum gravity effects will generically define $\frac{1}{M_P}$ corrections to the so defined QFT. 

As discussed in the previous section for a given de Sitter observer the corresponding QFT will be characterized by the algebra of local observables ${\cal{A}}_{dS}$ with support on the static patch that we have assumed is a type $III$ factor. In this case the commutant of this algebra corresponds to the algebra of local observables with support on the other static patch ( the one corresponding to the antipodal observer ). The natural time flow for this observer will be associated with the global time defined in the static patch and will have a generator that we have denoted $H_R$. As discussed in previous section and due to the infinite entanglement implicit in type $III$ factors $H_R$ is only well defined, as an operator in the QFT i.e. in the $N=\infty$ limit after it is appropriately regulated. In the former section we have briefly discussed the crossed product of ${\cal{A}}_{dS}$ with the time automorphism generated by the regulated $H_R$. However as it was also mentioned in case of de Sitter the observer can, in principle, use a different {\it emergent time}, namely the conformal time, that will foliate the whole planar patch. 

Let us intuitively equip the observer with a {\it clock} measuring conformal time $\eta$. At least at heuristic level this observer can use her {\it conformal clock} to register when a quantum fluctuation created in her static patch with comoving momentum $k$ {\it crosses the horizon}. In pure de Sitter this time is simply $\eta_k = \frac{1}{k}$. The quantity on which we are physically interested when doing inflation is essentially in evaluating the gravitational consequences, in the form of induced scalar curvature fluctuations, of the {\it variance} $\delta(\eta_k)$. Heuristically we can think of  $\delta(\eta_k)$ as defining the standard {\it time delay} at horizon crossing used in many approximated derivations of the power spectrum and to use this time fluctuation to induce, gravitationally, the scalar curvature fluctuation. Properly speaking in order to compare with CMB experiment, where what you actually measure is this variance at the moment the mode {\it reenters in the post inflationary epoch}, you need to assume that the amplitude of this variance is the same as the amplitude when the mode crosses the horizon during inflation. Thus and under this assumption a pure inflationary computation can be contrasted with the experimental CMB data. 

Let us, again very heuristically, think the amplitude during inflation of $\delta(\eta_k)$ as defining the different Fourier modes of the {\it conformal clock} uncertainty. The important physical question is how this amplitude depends on $k$ and on the underlying equation of state characterized by $\epsilon_{cl}$. You encode this information in the well known power spectrum of scalar curvature fluctuations. The key point underlying the former logic flow is that in the presence of gravity $\delta(\eta_k)$ induces pure quantum gravity fluctuations of the metric. Thus what you could naively think as an uncertainty in your (mode dependent) conformal clock becomes promoted, after coupling to gravity, into a real fluctuation of the metric. In the spirit of representing the cosmology in terms of the states in the crossed product Hilbert space, the former comment implies that the corresponding state cannot be of the type classical-quantum but instead must involve a quantum dependence of the state on the conformal time. We will discuss extensively this point later.

\subsubsection{Mukhanov-Sasaki variables}
In order to define the QFT for the scalar quantum fluctuations of the inflaton field on the planar patch the right gauge invariant variable to be used is the so called Mukhanov-Sasaki (MS) \cite{Mukhanov1} variable $v_k(\eta)$. The conformal time dependence of $v_k(\eta)$ is determined by the Chibisov-Mukhanov \cite{Mukhanov2} \cite{Mukhanov3} equation:
\begin{equation}\label{equation}
v^{''}_k + ( k^2 - \frac{z^{''}}{z}) v_k =0
\end{equation}
with $z= a\sqrt{\epsilon_{cl}}$ and $a(\eta)$ the classical conformal factor of the metric. This equation has a perfectly nice pure de Sitter limit $\epsilon_{cl} =0$ namely
\begin{equation}\label{dS}
v^{''}_k + ( k^2 - \frac{2}{\eta^2}) v_k =0
\end{equation}

Equation (\ref{equation}) has two very different regimes, namely the short wave regime with $k^2 >> \frac{z^{''}}{z}$ and the long wave regime with $k^2 <<  \frac{z^{''}}{z}$. The transient region corresponds to the time at which the mode with comoving momentum $k$ exits the horizon. A typical way to approximate the solution to (\ref{equation}) is by matching the solutions in both regimes at horizon exit. In the short wave regime the solution is of the form $A(k)e^{ik\eta}$ with $k$ playing the role of the energy conjugated to the conformal time. In the long wave regime the non decaying solution has the form $C(k)z$.

In the short wave regime the underlying quantum field theory is easy to understand, indeed in this regime  we can think in a standard quantum harmonic oscillator with frequency equal $k$. For this harmonic oscillator we can define the operator $\hat v_k$ in terms of two algebras of creation annihilation operators \footnote{See \cite{Martin}.} $a_{\pm k} a^{\dagger}_{\pm k}$ with $k\in R^{+}$ and to introduce a short wave Hamiltonian $\hat H_{sw} = \frac{k}{2} (a_{k} a^{\dagger}_k + a_{-k} a^{\dagger}_{-k})$ as well as a Fock representation of the creation annihilation algebras, characterized by a vacuum $|0\rangle$ satisfying $a_{\pm k}|0\rangle =0$. In these conditions $v_k(\eta)$ for $k^2 >> \frac{2}{\eta^2}$ can be represented as

\begin{equation}\label{quantum}
v_k(\eta) = \langle -k| e^{i \hat H_{sw} \eta} \hat v_k |0\rangle
\end{equation}
with $\hat v_k = \frac{1}{\sqrt{2k}} (a_k + a^{\dagger}_{-k})$ and $|k\rangle = a_k^{\dagger}|0\rangle$.
This representation leads to
\begin{equation}
v_k(\eta) = \frac{1}{\sqrt{2k}} e^{ik\eta}
\end{equation}
and $A(k) = \frac{1}{\sqrt{2k}}$. More precisely we can define the $\eta$ dependent operator
\begin{equation}
\hat v_k (\eta) = \frac{1}{\sqrt{2k}} (a_k(\eta) + a^{\dagger}_{-k}(\eta))
\end{equation}
In the short wave regime we can fix $a_k(\eta)$ by the Heisenberg equation
\begin{equation}\label{Heisenberg}
-i \frac{da_k}{d\eta} = [\hat H_{sw},a_k]
\end{equation}
that leads to $\hat v_k(\eta) |0\rangle= e^{ik\eta} \frac{1}{\sqrt{2k}} |k\rangle$ and to the representation
\begin{equation}\label{quantum2}
v_k(\eta) =  \langle -k| \hat v_k(\eta) |0\rangle
\end{equation}

The quantum representation (\ref{quantum2}) is a bit more complicated when we move into the long wave regime $k\eta << 1$. Formally we could define in this regime $a_k(\eta)$ by the analog of (\ref{Heisenberg}) with a different formal Hamiltonian $\hat H_{lw}$. Naively we could think $\hat H_{lw}$ as identical to $\hat H_{sw}$ but replacing the frequency $w=k$ in $\hat H_{sw}$ by the imaginary frequency $i\sqrt{\frac{z}{z^{''}}}$. This simple minded guess is however wrong. The correct answer is
\begin{equation}
\hat H_{lw} = -i\frac{z'}{z}(a_ka_{-k} - a^{\dagger}_k a^{\dagger}_{-k})
\end{equation}
Indeed defining $\hat H= \hat H_{sw} + \hat H_{lw}$ and 
\begin{equation}
\frac{d \hat v_k}{d\eta} = [\hat H, \hat v_k]
\end{equation}
we easily get the operator version of equation (\ref{equation}) namely
\begin{equation}
\hat v_k(\eta)^{''} + (k^2-\frac{z^{''}}{z}) \hat v_k(\eta) =0
\end{equation}
where $\frac{z^{''}}{z}$ is a c-number function.

Note that while $[H_{sw},\hat v_k]$ is proportional to $\hat v_k$ this is not the case for $[\hat H_{lw}, \hat v_k]$. 

\subsubsection{A comment on the Spectral index}
In its simplest version we can introduce a {\it classical} quasi de Sitter modification of equation (\ref{dS}) replacing $\frac{2}{\eta^2}$ by $\frac{\beta(\beta+1)}{\eta^2}$ (with $\beta =-2-\delta$ and $\delta=0$ corresponding to the pure dS limit $\beta=-2$). For this quasi de Sitter equation we can easily find $C(k)$ using the matching at horizon exit with the result
\begin{equation}
v_k(\eta) = \frac{1}{k^{3/2}} (k\eta)^{\delta}
\end{equation}
So the power spectrum defined as $k^3|v_k(\eta)|^2$ scales as $(k\eta)^{2\delta}$ that leads to the spectral index
\begin{equation}
(1-n_s) = 2\delta
\end{equation}

This simple exercise shows that the quasi de Sitter modification of the classical equation defining $v_k(\eta)$ leads to a well defined deviation from scale invariance and to a well defined spectral index. 


\subsubsection{The scalar curvature fluctuations}

In order to make contact with CMB experiment we need the power spectrum of some variable representing $\frac{\delta T}{T}$ with $T$ the average background temperature. Making use of Sachs-Wolfe effect this quantity is directly related with the gravitational potential $\Phi$. Thus we are now interested in the pure quantum gravity fluctuation of the metric induced by the quantum fluctuation of the inflaton field. In terms of the MS variable this quantity is defined as
\begin{equation}
\zeta_k(\eta) = \frac{v_k(\eta)}{a \sqrt{2\epsilon_{cl}} M_P}
\end{equation}
 The relation between $\zeta_k(\eta)$ and the gravitational potential follows from the classical Einstein equations, namely
\begin{equation}
\zeta = \frac{\frac{\Phi'}{\cal{H}} + \Phi}{\epsilon_{cl}} + \Phi
\end{equation}
with ${\cal{H_b}} = \frac{a'}{a}$. 

 The power spectrum for $\zeta_k$ is then defined as
\begin{equation}
{\cal{P}}_{\zeta}(k,\eta) = k^3|\zeta_k(\eta)|^2
\end{equation}
 Let us now evaluate this quantity at some pivot scale $k_0$ and at horizon exit time $\eta_0$. This time should be evaluated using the quasi de Sitter metric i.e. $a(\eta_0)= \frac{k_0}{H(\eta_0)}$ defined by the corresponding inflaton potential. Using the former definitions we easily get
\begin{equation}\label{qdsc}
\zeta_{k_0}(\eta_0) = \frac{1}{k_0^{3/2}} \frac{(k_0\eta_0)^{\delta}}{\sqrt{2\epsilon(\eta_0)} M_P a(\eta_0)\eta_0}
\end{equation}
and for the curvature power spectrum
\begin{equation}
{\cal{P}}(k_0,\eta_0) = k_0^3 |\zeta_k(\eta_0)|^2 = \frac{(k_0\eta_0)^{2\delta}}{2\epsilon(\eta_0) a^2(\eta_0) \eta_0^2 M_P^2} \sim \frac{(k_0\eta_0)^{2\delta} H^2(\eta_0)}{2\epsilon(\eta_0) M_P^2}
\end{equation}
In this expressions $\delta$ is the spectral index defined at $\eta_0$ i.e. in terms of the slow roll parameters evaluated at $\eta_0$. When we change the pivot scale $k_0$ to some close $k$ we get, at first order, the renormalization group equation
\begin{equation}
{\cal{P}}(k) = {\cal{P}}(k_0) (\frac{k_0}{k})^{2\delta}
\end{equation}
In the {\it time delay} approximation we try to represent $k_0^{3/2}\zeta_{k_0}(\eta_0)$ as $H(\eta_0) \delta \eta(k_0)$ with
\begin{equation} 
\delta \eta(k_0) = \frac{(k_0\eta_0)^{\delta}}{\sqrt{2\epsilon} M_P }
\end{equation}

\subsection{Back to the crossed product: Toward a predictive scheme for inflation}
Generically horizons and in particular cosmological horizons lead to type $III$ factors and consequently to a natural way to implement time translations in terms of {\it outer} automorphisms. As stressed above the key physics goal of crossed products is to extend the algebra of local observables in such a way that time translations become represented by an inner automorphism. In more plain words that the Hamiltonian, as a well defined operator, can be expressed as a polynomial in the local observables. As already stressed this extended algebra is defined in terms of "paths" i.e. continuous and bounded functions from "time" into the algebra of local observables and the Hilbert space representation is defined using equally continuous and square integrable functions on the GNS Hilbert space used to define the representation of our starting type $III$ factor.

In inflationary Cosmology we are working in the planar patch and we use as time the conformal time $\eta$. Using Kruskal coordinates we can define incoming and outgoing modes and to use the Heisenberg algebra representation of these modes to define the algebras ${\cal{A}}^{\pm}$. Working with conformal time naturally forces us to work with {\it "Bogolyubov paths"} $a(\eta)$  valued in the algebra ${\cal{A}}^{+} \otimes {\cal{A}}^{-}$. These paths, as described in the previous section are generated by some formal Hamiltonian through standard Heisenberg equations. However this formal Hamiltonian is self consistently  written in terms of the paths $a(\eta)$.

With these ingredients at hand it is natural to make the formal guess that these functions from $\eta$ into the algebra are indeed elements of some underlying crossed product and that the states depending on $\eta$ are equally playing the deep role of defining a representation of a crossed product. At this point the skeptic reader can certainly ask: Why? Although we don't have a compelling answer the heuristic reason can be that in this case the transformation relating the creation annihilation operators at different times should be defined by an {\it outer automorphism}. 

In the case of inflationary Cosmology it looks natural to focus separately on each mode of comoving momentum $k$. In these conditions a vary natural "path-state" could be the state $|k,\eta\rangle$ defined in (..) where we define this state using the {\it pure de Sitter} definition. This is not a simplifying hypothesis but part of our philosophy to look for states in $L^2(R^+, {\cal{H}}^{dS}_{GNS})$ where we keep fixed the pure dS Hilbert space. The state $|k,\eta\rangle$ represents, in the purification defined by the algebras ${\cal{A}}^{\pm}$, a very entangled state between $\pm$ modes. From the point of view of the underlying crossed product we can think of the operator $\hat X$ as such that $e^{i\eta\hat X} |k,\eta_0\rangle = |k,\eta_0+\eta\rangle$. Certainly this state is not eigenvector of $\hat X$ although it represents the effective vacuum for modes $k$ at time $\eta$. It is this non trivial transformation of $|k,\eta\rangle$ under $\hat X$ what induces non trivial {\it phases} in the definition of $|k,\eta\rangle$, namely

\begin{equation}\label{state}
|k,\eta \rangle = \sum_n c((k\eta),n) e^{in\Phi((k\eta))}|n_k,n_{-k}\rangle
\end{equation}
where we sum over all integers and  
\begin{equation}
c((k\eta),n)=C \tanh({r(k\eta)})^{n}
\end{equation} 
 with $C$ a normalization constant, 
 \begin{equation}
 r(k \eta) = - \sinh^{-1}(\frac{1}{2k\eta})
\end{equation} 
and with the phase given by
\begin{equation}
\Phi(k\eta) = -\frac{\pi}{4} -\frac{1}{2} \tan^{-1}(\frac{1}{2k \eta})
\end{equation}

The quantum variance $\Delta( \hat X^2)$ is defined as the variance of the phase in (\ref{state}) namely \cite{GJ1},\cite{GJ2} \footnote{Note that in our present context we cannot use the analog of the replica method to evaluate this quantum variance.}
\begin{equation}
\Delta( \hat X^2 (k\eta)) = ( \sum_n c(k\eta,n)^2 (\frac{\partial \Phi(k\eta,n)}{\partial (k\eta)})^2 - (\sum_n c(k\eta,n)^2 (\frac{\partial \Phi(k\eta,n)}{\partial (k\eta)}))^2)
\end{equation}

For pure de Sitter the former expression can be numerically evaluated introducing a upper cutoff in the sum over $n$ with the result
\begin{equation}
\Delta( \hat X^2)(k \eta) = \frac{1}{4 (k\eta)^8} (k\eta)^{\alpha_F(k\eta)}
\end{equation}
where the quantum tilt $\alpha_F(k\eta)$ has a very precise dependence on $(k\eta)$. In previous publications \cite{GJ1},\cite{GJ2} we have evaluated the tilt $\alpha_F(k\eta)$ and study the sensibility of the numerical analysis on the cutoff over $n$. For future convenience let us define $\hat \Delta(\hat X^2)(k\eta) =: (k\eta)^6 \Delta( \hat X^2)(k \eta) = \frac{1}{(k\eta)^2} (k\eta)^{\alpha_F(k\eta)}$. 

Once we reach this point we are in good conditions to transform the qualitative relation (\ref{keys}) into a precise definition of scalar curvature fluctuations. Indeed the scalar curvature fluctuations should be defined by {\it the gravitational potential $\Phi$ induced by the variance $\hat \Delta(\hat X^2)(k\eta)$.}

The gravitational potential created by the variance $\hat \Delta(\hat X^2)(k\eta)$ is given by up to constants order one by
\begin{equation}
\Phi = \sqrt{\hat \Delta(\hat X^2)(k\eta)} \frac{H_b}{M_P}
\end{equation}
that leads to the basic definition of the scalar curvature fluctuations in terms of the potential $\Phi$, again up to factors order one, as
\begin{equation}\label{master}
\zeta_k(\eta) = \sqrt{\hat \Delta(\hat X^2)(k\eta)} \frac{H_b}{M_P}
\end{equation}
The conceptual reading of the former equation would be that:

{\it The curvature scalar perturbations define, for each mode, the gravitational potential induced by the variance of $\hat X$ in the quantum state $|k\eta\rangle$ formally living in $L^2(R^+,{\cal{H}}_{dS})$.}

Obviously relation (\ref{master}) opens the possibility to make model in dependent predictions about the standard inflation phenomenological parameters $(1-n_s)$ as well as $\epsilon$. To extract these predictions what we need to do is to compare the $\zeta_k$ defined by (\ref{master}) with the expressions derived in the quasi de Sitter approximation (\ref{qdsc}).

\subsection{Numerical predictions}
In order to get some numerical predictions on the inflationary parameters using the former scheme we will use two matching conditions. The first is basically the one derived from (\ref{master}) evaluated at horizon exit. The second matching condition puts in correspondence the quasi de Sitter deviation of $\omega^2(k,\eta)$ with $\omega^2(k,\eta) = k^2-\frac{\beta.(\beta-1)}{\eta^2}$ due to the quasi de Sitter deformation and the quantum contribution to $\omega^2(k,\eta)$ defined in terms of $\Delta(\hat X^2)$.

Let us start with the second matching condition, already discussed in \cite{GJ}. We can define $\Delta(\hat X) (k,\eta) = \frac{\delta k}{k} (\eta)$ and the quantum contribution to $\frac{\omega^2(k,\eta)}{a^2}$ as
\begin{equation}
\frac{k \delta k}{a^2} = \frac{(k\eta)^{1/2 \alpha_F(k\eta)}}{2\eta^2 a^2}
\end{equation}
This quantity should be matched with the quasi de Sitter expression, namely
\begin{equation}
\frac{1}{a_{qdS}^2 \eta^2}(3\delta +\delta^2)
\end{equation}
In order to define $a_{ads}$ we use a pivot scale $k_0$ and define $a_{ads}= \frac{1}{H_0 \eta} (k_0\eta)^{\delta}$ fixing the pivot Hubble as $H_0=H(\eta=\frac{1}{k_0})$. In these conditions the former matching condition becomes
\begin{equation}
(k_0\eta)^{\frac{\alpha_F(k_0\eta)}{2}} (k_0\eta) H_0^2 = (6\delta +2\delta^2) H_0^2 (k_0\eta)^{2\delta}
\end{equation}
that becomes the following equation for $\delta$
\begin{equation}
\alpha_F(6\delta +2\delta^2) = 4\delta
\end{equation}
Using the numerical value of $\alpha_F(k\eta)$ set by $\Delta(\hat X^2)$ we get the solution $\delta$ and the corresponding spectral index $(1-n_s)=2\delta$ as
\begin{equation}
(1-n_s) = 0.0328
\end{equation}
Once we have identified $\delta$ we can use the second matching condition
(\ref{master}) at horizon exit. This yields
\begin{equation}
(k_0\eta_0) \frac{H}{\sqrt{2\epsilon}M_P} = \frac{(k_0\eta_0)^{\frac{\alpha_F(k_0\eta_0)}{2}} H}{(k_0\eta_0) M_P}
\end{equation}
This matching condition set the value of $\eta_0$ by the condition
\begin{equation}
\alpha_F(k_0\eta_0) = 2\delta
\end{equation}
and fix the value of $\epsilon$ as
\begin{equation}
\sqrt{2 \epsilon} = (k_0\eta_0)
\end{equation}
Using now the value of $\delta$ derived from the first matching condition we get 
\begin{equation}\label{slowroll}
\epsilon =0.0027
\end{equation}
Note that these predictions that use, as unique input, the value of the quantum tilt $\alpha_F(k\eta)$ that we derive directly from $\Delta(\hat X^2)$ on the state $|k,\eta\rangle$, are in good agreement with the results from Planck \cite{Planck}. In summary these are the numerical predictions implied by representing the scalar curvature fluctuations as the gravitational effect of the quantum variance using (\ref{master}).

\subsubsection{A comment on causal diamonds}
In \cite{Verlinde1},\cite{Verlinde2},\cite{banks} causal diamonds on the boundary in AdS have been considered. Using the Ryu-Takayanagi formula these causal diamonds are associated with an entropy $\frac{A(\Sigma)}{4G}$ and with a modular Hamiltonian $K$ with $\langle K\rangle =\frac{A(\Sigma)}{4G}$. Using the replica method you evaluate $\Delta(K^2)$ and the fluctuations of the gravitational potential $\Phi$ induced by these fluctuations at the horizon. Conceptually this is similar to the former derivation of scalar curvature fluctuations at horizon exit in the inflationary context. The result obtained in \cite{Verlinde2} using the replica method in AdS is the analog of what you will get for scalar curvature fluctuations at the horizon for $\epsilon=1$. As a very speculative comment we could think that if our former analysis of scalar curvature fluctuations in inflation extends to the experimental setup used in \cite{Verlinde2}, then we should expect an enhancement of the causal diamond fluctuations by a factor $\frac{1}{\sqrt{\epsilon}}$ for some effective $\epsilon$ likely of the order defined in (\ref{slowroll}).

\acknowledgments
 I thank, R. Jimenez and S.Das for discussions and T.Banks for bringing to my attention the results on fluctuations in causal diamonds. This work was supported by grants SEV-2016-0597, FPA2015-65480-P and PGC2018-095976-B-C21.

\end{document}